\documentclass[aps,
prb,
reprint,
preprintnumbers,
superscriptaddress,
amsfonts,
amssymb,
amsmath
]{revtex4-2}
\usepackage{bm,latexsym,mathrsfs,enumerate,bigints}
\usepackage{upgreek}
\usepackage[table,x11names]{xcolor}
\usepackage[breaklinks=true,
unicode=true,
urlcolor = RoyalBlue4,
colorlinks = true,
citecolor = RoyalBlue4,
linkcolor = blue
]{hyperref}

\usepackage{graphicx}
\graphicspath{{../figs},{./figs/}}
\usepackage{mathtools}
\usepackage{physics}
\usepackage[cal=boondoxo,frak=boondox,scr=rsfs]{mathalfa}
\usepackage{savesym}
\usepackage[most]{tcolorbox}
\savesymbol{comment}
\usepackage{easyReview}

\usepackage{gitinfo2}
\renewcommand{\vec}[1]{\bm{#1}}

\usepackage{orcidlink}

\begin{document}

\title{Manipulation by magnetic frustration in ferrotoroidal spin chains via curvature and torsion}

\author{Oleksandr~V.~Pylypovskyi \orcidlink{0000-0002-5947-9760}}
\email{o.pylypovskyi@hzdr.de}
\affiliation{Helmholtz-Zentrum Dresden-Rossendorf e.V., Institute of Ion Beam Physics and Materials Research, 01328 Dresden, Germany}
\affiliation{Kyiv Academic University, Kyiv 03142, Ukraine}

\author{Enrico~Di~Benedetto\,\orcidlink{0009-0003-3613-5257}}
\affiliation{Helmholtz-Zentrum Dresden-Rossendorf e.V., Institute of Ion Beam Physics and Materials Research, 01328 Dresden, Germany}
\affiliation{Dipartimento di Fisica e Chimica ``E. Segr\`{e}'', Universit\`{a} degli Studi di Palermo, 90123 Palermo, Italy}

\author{Carmine~Ortix \orcidlink{0000-0002-6334-0569}}
\affiliation{Dipartimento di Fisica ``E. R. Caianiello'', Universit\`{a} di Salerno, IT-84084 Fisciano (SA), Italy}

\author{Denys~Makarov \orcidlink{0000-0002-7177-4308}}
\email{d.makarov@hzdr.de}
\affiliation{Helmholtz-Zentrum Dresden-Rossendorf e.V., Institute of Ion Beam Physics and Materials Research, 01328 Dresden, Germany}

\date{August 30, 2024}

\begin{abstract}
Geometric effects in curvilinear nanomagnets can enable chiral, anisotropic and even magnetoelectric responses. Here, we study the effects of magnetic frustration in curvilinear (quasi-)1D magnets represented by spin chains arranged along closed space curves of constant torsion. Considering the cases of easy- and hard-axis anisotropy in ferro- and antiferromagnetic samples, we determine their ground states and analyze the related  magnetoelectric multipoles. A constant torsion along the chain results in alternating regions of high and low curvature, facilitating the spin spiral state perturbed by the (anti)periodic boundary conditions on the magnetic order parameter. While easy-axis ferromagnetic chains develop a purely toroidal configuration with the magnetic toroidal moment oriented along the geometry symmetry axis, hard-axis antiferromagnetic chains support multiple magnetic toroidal domains. Our findings suggest that tailoring curvature and torsion of a spin chain enables a new physical mechanism for magnetic frustration, which can be observed in the inhomogeneity of the magnetic order parameter and in the local ferrotoroidic responses.
\end{abstract}

\maketitle

\section{Introduction}

Non-trivial geometry of low-dimensional objects like wires, ribbons and shells emerged as an effective way to modify their responses to external stimuli~\cite{Gentile22}. These phenomena are intensively explored for a broad range of condensed matter systems, including two-dimensional materials~\cite{Vozmediano10} and topological insulators and semimetals~\cite{Grushin16} in addition to more established curvilinear semiconductors~\cite{Gentile22}, superconductors~\cite{Fomin22,Makarov22} and magnetic materials~\cite{Raftrey22,Makarov22a,Sheka22}. Very recently, the geometry of curvilinear magnetic wires with intrinsic Dzyaloshinskii–Moriya interaction (DMI) was explored to realize artificial systems characterized by magnetoelectric multipoles~\cite{Ortix23}. Indeed, geometric effects  can result in a magnetically induced breaking of inversion symmetry and consequently to the appearance of magnetoelectic monopoles, toroidal and quadrupole moments. 
Materials with finite magnetoelectric multipoles moments are of major fundamental interest in the study of magnetoelectric coupling phenomena~\cite{Fiebig16}. Typically, magnetoelectric momopoles, toroidal and quadrupolar moments in materials are realized at the level of crystal structure. The use of geometry induced breaking of inversion symmetry can extend the number of magnetoelectric materials, since it can virtually equip any appropriately curved magnetic material with higher order magnetoelectric multipoles. 

Here, we study magnetic states in spin chains following space curves of constant torsion. For completeness, we address the cases of ferro- and antiferromagnetic coupling in the spin chain. The geometries studied here belong to the family of spherical epicycles, which are characterized by the 3- and 4-fold symmetry axis only with the 3 and 4 number of knots without self-intersections, respectively. These systems highlight the role of a finite torsion on the physical properties of 3D nanoarchitectures. We describe the ground states in samples with easy- and hard-axis magnetic anisotropy including the case of odd number of spins in the chain, which leads to magnetic frustration. The equilibrium magnetic textures are determined by the competition between the strength of anisotropy, geometry-driven DMI and (anti)periodic boundary conditions. The latter depends on the number of knots and spins in case of the AFM exchange coupling. We show that in this case the magnetic frustration of the ground state (the total winding phase of the order parameter is an odd or even multiple of $\pi$) is dependent not only on the number of spins, but also on the number of knots in the geometry. The magnetic texture in epicycles breaks the space-inversion symmetry and allows for higher-order magnetoelectric multipoles. In particular, while spin chains with ferromagnetic coupling and easy axis anisotropy are characterized by a macrotoroidal moment, hard-axis systems can be split into multiple toroidal domains. 

The paper is organized as follows. In Sec.~\ref{sec:curveffect} we introduce curvature effects in magnetic systems. In Sec.~\ref{sec:model}, the geometry, spin-lattice model and the respective nonlinear $\sigma$-model are introduced. In Sec.~\ref{sec:magnetic-states}, we describe ground states of closed ferromagnetic and antiferromagnetic spin chains analyzing cases of the easy-, hard-axis anisotropy, magnetically soft chains with dominating dipolar interaction and chains with 
antiferromagnetic exchange and odd number of spins. In Sec.~\ref{sec:ferrotoroidal}, the static ferrotoroidal ordering in these systems is described. In Sec.~\ref{sec:discussion}, we summarize the identified magnetic states and describe the interplay between magnetic frustration originating either purely from the sample geometry (distribution of curvature and torsion) or due to the odd number of spins. In Appendix~\ref{app:geometry}, details of the construction of the epicycle geometry are provided. Appendix~\ref{app:frustrated} describes boundary conditions in the $\sigma$-model for chains with antiferromagnetic exchange and odd number of spins. In Appendices~\ref{app:eq-states} and~\ref{app:simulations}, we provide the equations of state and describe spin-lattice simulations, respectively.

\section{Curvature effects in magnetism}
\label{sec:curveffect}

In magnetism, the static effects of geometry are usually related to the presence of geometry-tracking interactions like magnetostatics, interfacial phenomena or DMI. Then, breaking of space-inversion symmetry due to the curved shape of the sample is mapped onto the behavior of the magnetic vector order parameter. 

A strong attention has been paid to high symmetry samples, with a particular focus on (quasi) 1D systems~\cite{Sheka22}. Thin bent nanowires or spin chains with ferro- or antiferromagnetic exchange are convenient objects of study that capture the hallmarks of more complex geometries. Such chains are characterized by curvature $\kappa$ and torsion $\tau$, two geometric quantities describing in-plane bends and screw deformation of the given space curve respectively. Prototypical examples of such systems are represented by helices, characterized by constant $\kappa$ and $\tau$, which can be readily realized in experiments~\cite{Phatak14,Magdanz17,Nam18,Sanz-Hernandez20,Phatak20,Fullerton24}. On one hand, thin helical ferromagnetic (FM) nanowires have the ground state tilted from the tangential direction due to the geometry-driven anisotropy stemming from exchange interaction~\cite{Sheka15a,Pylypovskyi16} and support chiral domain walls~\cite{Pylypovskyi16}. The selection of $\kappa$ and $\tau$ can be used for modulating the magnetic states in intrinsically chiral materials~\cite{Volkov18} or tailoring the properties of heterostructures with piezoelectric~\cite{Volkov19b} or superconducting matrices~\cite{Salamone24a}. On the other hand, antiferromagnetic (AFM) helical systems show a broad range of transport phenomena, which can be tailored by geometric parameters~\cite{DasGupta22,Mondal23,DasGupta23}. At present, the analysis of geometric effects is mainly limited to planar geometries ($\tau = 0$), where curvature gradients act as pinning potentials for the non-collinear FM and AFM magnetic textures~\cite{Yershov16,Moreno17,Volkov19c,Yershov22a,Bittencourt24a}. In geometries following space curves, the gradients of $\kappa$ and $\tau$ in 3D nanostructures result in the self-propelling of domain walls~\cite{Yershov18,Bittencourt22,Zhao23,Askey24}.

Inversion symmetry breaking in curvilinear magnetic systems can lead to a broader range of phenomena, such as geometry-induced magnetoelectricity due to the possibility to support non-vanishing magnetoelectric multipoles~\cite{Ortix23}. This fundamental knowledge is of relevance for the analysis of the response of a broad class of single-molecule magnets in a toroidal ground state~\cite{Bogani08,Ungur14,Murray22}.

In addition to extensively discussed open-end systems like helices, it is insightful to focus on closed space curves, as they allow discussing effects in terms of the topological properties of the sample geometry. In this respect, the correlation between the topology of the geometry and the topology of the magnetic texture in effectively 1D systems can be established~\cite{Pylypovskyi15b}. 

Furthermore, much attention is usually dedicated to the effects of curvature, while torsion is usually considered as a secondary parameter. In strictly 1D systems, torsion reveals itself only in the presence of a finite curvature~\cite{Gaididei17}. Hence, the impact of torsion on the behavior of curvilinear magnetic systems is not well studied yet. This seems to be a severe limitation, especially considering that the case of nonzero torsion distinguishes a planar curved wire from a truly 3D curved wire. To facilitate this understanding, in this manuscript we focus on closed space curves with constant torsion. 

\section{The model}
\label{sec:model}

\subsection{Geometry}

\begin{figure*}[t]
	\centering
	\includegraphics[width=\linewidth]{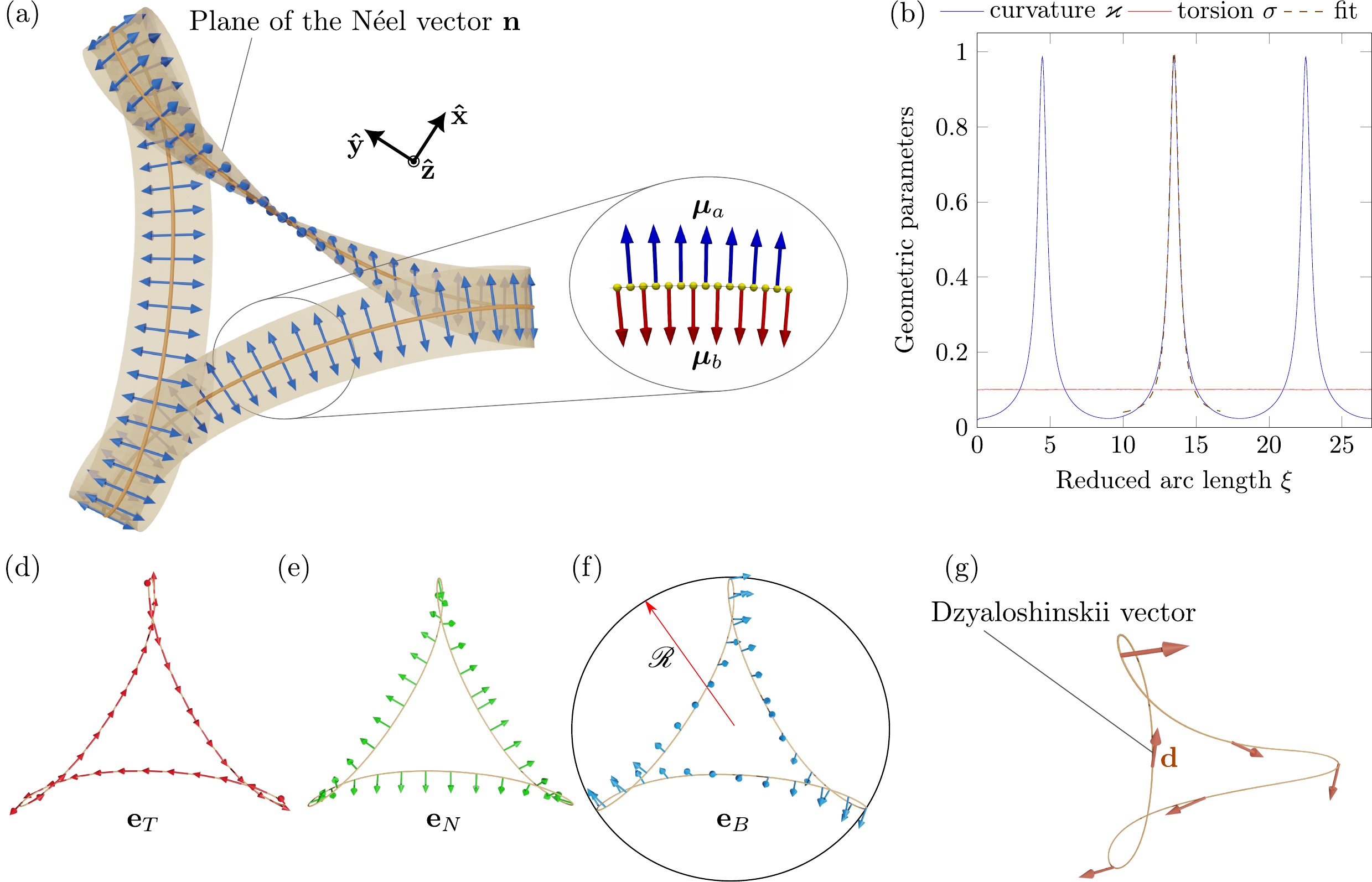}
	\caption{
		(a)~An antiferromagnetic (AFM) spin chain whose shape follows the epicycle $\vec{\gamma}_{-3}^1$. The direction of the N\'{e}el vector $\vec{n}$ for each dimer is shown in blue, whereas the envelope surface resulting from the continuous variation of the $\vec{n}$ vector along the curve is drawn in light yellow. In the inset, schematics of the two magnetic sublattices $\vec{\mu}_a$ and $\vec{\mu}_b$ is shown. 
		(b)~Reduced curvature $\varkappa$ for torsion $\sigma = 0.1$. Each peak of the dimensionless curvature $\kappa$ can be well fitted with a Lorentzian profile, shown for comparison with an orange dashed line.
		(d--f)~TNB frame in real space for the geometry shown in panel (a). The red arrow indicates the radius of the circumscribed ring $\mathscr{R}$. 
		(g)~The direction of the Dzyaloshinskii vector $\vec{d}$ in some representative points along the epicycle. Arrows are not to scale.		
	}
	\label{fig:first}
\end{figure*}

We consider spin chains arranged along a space curve $\vec{\gamma}(s) \in \mathcal{C}^3(\mathbb{R})$ characterized by a constant torsion $\tau = [(\partial_s\vec{\gamma} \times \partial_s^2\vec{\gamma})\cdot \partial_s^3\vec{\gamma}]/|\partial_s^2\vec{\gamma}|^2$, where $s$ is the arc length [Fig.~\ref{fig:first}(a)]. 
Exemplary mathematical approaches to design such geometries include B\"{a}cklund transformation~\cite{Calini98}, building curves dual to the curves of constant curvature $\kappa = |\partial_s\vec{\gamma}\times \partial_s^2\vec{\gamma}|/|\partial_s\vec{\gamma}|^3$~\cite{Monterde09} and transformation of spherical curves $\vec{\beta}$~\cite{Koenigs1887}
\begin{equation}
	\label{eq:koenigs}
	\vec{\gamma}(s) = \frac{1}{\tau} \int_0^s \left(\vec{\beta} \times \dfrac{d\vec{\beta}}{d\tilde{s}}\right) d \tilde{s}.
\end{equation}
Here, $\vec{\beta}$ plays a role of the binormal vector $\vec{e}_\textsc{b}$ for~$\vec{\gamma}$. It has positive geodesic curvature and possesses zero Peano direction~\cite{Weiner77}. The latter implies the presence of an inversion center for certain projections of the curve. In this work, we use the approach~\eqref{eq:koenigs} and generate spherical curves following the method by \citet{Bates13}. We chose two-parametric epicycles $\vec{\beta}_p^r$, where $r$ counts the number of traces of a unit sphere following the epicycle, and $p$ is a winding number equal by absolute value to the number of knots on the curve, see Appendix~\ref{app:geometry} for details. The associated curve $\vec{\gamma}_p^r$ inherits the number of knots $|p|$. The transformation~\eqref{eq:koenigs} lifts the presence of self-intersections, see example of $\vec{\gamma}_{-3}^1$, which is superimposed with an AFM texture in Fig.~\ref{fig:first}(a). The resulting symmetry of the curve is the $|p|$-fold rotation axis $C_{|p|}$. In the following, we introduce the laboratory Cartesian reference frame in such a way to have $\vec{\hat{z}}$ axis along the symmetry axis of the epicycle.

As an example, we consider the case of $\vec{\gamma}_{-3}^1$. According to its $C_3$ symmetry, the curvature of $\vec{\gamma}_{-3}^1$ exhibits three peaks corresponding to the number of knots of the generating curve [Fig.~\ref{fig:first}(b)]. The shape of $\kappa(s)$ in the vicinity of extrema can be well approximated by a Cauchy--Lorentz function (Appendix~\ref{app:geometry}). This is valid for any $\vec{\gamma}^{1}_{p}$.

Following the Frenet--Serret approach, the local reference frame on $\vec{\gamma}$ is determined by the tangential (T) $\vec{e}_\textsc{t} = \partial_s\vec{\gamma}$, normal (N) $\vec{e}_\textsc{n} = (\partial_s^2\vec{\gamma})/\kappa$ and binormal (B) $\vec{e}_\textsc{b} = \vec{e}_\textsc{t} \times \vec{e}_\textsc{n}$ directions, respectively [Fig.~\ref{fig:first}(d--f)]. The direction of $\vec{e}_\textsc{b}$ shown in Fig.~\ref{fig:first}(f) has clear evidence of the broken symmetry at each knot. A curve generated by Eq.~\eqref{eq:koenigs} can be circumscribed in a circle of radius $\mathscr{R} = C/\tau$ with a constant $C$ being dependent on $\vec{\beta}$ [Fig.~\ref{fig:first}(f,g)]. In contrast to plane rings whose characteristic size is determined by $\kappa$, the characteristic size $\mathscr{R}$ of $\vec{\gamma}$ grows with smaller torsion. In the following, we focus on the curves $\vec{\gamma}_{-3}^1$ and $\vec{\gamma}_{-4}^1$.

\subsection{Spin-lattice Hamiltonian}

In the magnetic Hamiltonian of the spin chain, we take into account the nearest-neighbor exchange, the single-ion anisotropy and the dipolar interaction, writing
\begin{equation}
	\label{eq:ham}
	\begin{split}
		\mathscr{H} = & -\dfrac{JS^2}{2} \sum_i \vec{\mu}_i \cdot \vec{\mu}_{i+1} - \dfrac{g\mu_\textsc{b}S}{2} \sum_i \vec{\mu}_i \cdot \vec{H}_\text{d}\\
		& - \dfrac{\mathcal{K}S^2}{2}\sum_i (\vec{\mu}_i\cdot \vec{e}_\textsc{t}^i)^2,
	\end{split}
\end{equation}
where $J$ is the exchange integral, $S$ is the length of the spin, $\vec{\mu}_i$ is the unit vector of $i$-th magnetic moment, $g = 2$ is the Land\'{e} factor, $\mu_\textsc{b}$ is the Bohr magneton. The dipolar field reads $ \vec{H}_\text{d} = - g\mu_\textsc{b}S \sum_{j \neq i} \left[ \vec{\mu}_j/r_{ij}^3 - 3 \vec{r}_{ij}(\vec{\mu}_j \cdot \vec{r}_{ij})/r_{ij}^5 \right]$ with $\vec{r}_{ij}$ being the distance between $i$-th and $j$-th magnetic moments. The last term in~\eqref{eq:ham} represents the single-ion anisotropy with the coefficient $\mathcal{K}$ and anisotropy axis $\vec{e}_\textsc{t}^i \equiv \vec{e}_\textsc{t}(s_i)$, where $s_i$ is the arc length for the $i$-th spin. Here, the sums run over all $N$ spins of the chain and $(N+1)$-th spin corresponds to the first one. The distance between neighboring spins is $a$, which gives the curve length $L = (N+1)a$. 

It is convenient to describe the macroscopic state of both systems, with ferromagnetic (FM) and antiferromagnetic (AFM) exchange coupling ($J > 0$ or $J < 0$, respectively), in terms of the vector order parameter $\vec{n}$. 
For the case of $J > 0$, the magnetic state of the chain is characterized by the unit vector of the order parameter $\vec{n}_i \equiv \vec{\mu}_i$, which is associated with the direction of the local magnetic moment of the chain. We refer to these samples as \emph{FM epicycles}.

For the case of $J < 0$, we consider cases of even and odd values of $N$. For both of them, the primary order parameter $\vec{n}$ is the N\'{e}el vector (staggered magnetization), supplemented by the vector of ferromagnetism $\vec{m}$. In spin chains, these vectors can be introduced as $\vec{n}_i = (\vec{\mu}_{2i-1}-\vec{\mu}_{2i})/2$ and $\vec{m}_i = (\vec{\mu}_{2i-1}+\vec{\mu}_{2i})/2$ with $i=\overline{1,N/2}$. They obey the relations $\vec{n}_i\cdot \vec{m}_i = 0$ and $n_i^2 + m_i^2 = 1$ for each $i$. With this definition, $\vec{n}_i$ and $-\vec{n}_i$ represent the same physical state of the lattice indicated by the double arrows in Fig.~\ref{fig:first}(a), where the magnetic sublattices are labeled by indices $a,b$. These samples are referred to as \emph{AFM epicycles}. The same procedure can be carried out for the chains with odd $N$ if the last magnetic moment, labeled as $\vec{\mu}_0$, is excluded from this dimerization procedure. These samples are referred to as \emph{frustrated epicycles}. 

\subsection{Representation of energy in the continuous limit}
\label{sec:sigma-model}

It is convenient to replace summation over spin chain by integration over the continuous counterparts of the magnetic vectors, $\vec{n}_i \to \vec{n}(\vec{r})$, and, in case of $J < 0$, also $\vec{m}_i \to \vec{m}(\vec{r})$. The anisotropy energy with the axis along $\vec{e}_\textsc{t}$ reads $E_\text{a}[\vec{n}] = -K \int n_\textsc{t}^2 ds$. The anisotropy coefficient $K = K_\text{a} + K_\text{dip}$ consists of the contribution from the single-ion anisotropy $K_\text{a} = \mathcal{K}S^2/(2a)$~\cite{Pylypovskyi21e} and dipolar interaction $K_\text{dip}$. The latter constant is characterized by the shape anisotropy and depends on the sign of the exchange integral, $K_\text{dip}^\text{fm} \approx 3.6 (g\mu_\textsc{b}S)^2/a^4$ for $J > 0$ (effective easy-axis) and $K_\text{dip}^\text{afm} \approx -2.7(g\mu_\textsc{b}S)^2/a^4$ for $J < 0$ (effective hard-axis)~\cite{Pylypovskyi20}. 

The exchange energy for the AFM spin chain with an even number of spins $N$ reads~\cite{Pylypovskyi21e}
\begin{equation}\label{eq:exchange}
	E_\text{x}[\vec{m},\vec{n}] = \int \left[ \Lambda m^2 + A_0(\partial_s\vec{n})^2 + \lambda \vec{m}\cdot (\partial_s\vec{n})\right]ds
\end{equation}
with $\Lambda = 2|J|S^2/a$ being the constant of the uniform exchange, $A_0 = |J|S^2a$ the exchange stiffness and $\lambda = 2|J|S^2$ the parity-breaking coefficient. If the magnetization is small, it can be excluded from the energy functional as a driving variable with the respective rescaling of the expression~\eqref{eq:exchange} as $E_\text{x}[\vec{n}] = A \int \vec{n}'^2 ds$ with $A = A_0/2$. In this case, $\vec{m} = - 0.5a \partial_s\vec{n}$. The same expression $E_\text{x}[\vec{n}]$ is valid for FM chains. The closure of $\vec{\gamma}$ here can be taken into account via the periodic boundary conditions on the primary order parameter, $\vec{n}(0) \equiv \vec{n}(L)$~\cite{Pylypovskyi21e,Borysenko22}. 

For the case of frustrated chains, the derivation procedure of the expression for exchange energy in the form~\eqref{eq:exchange} brings about two terms corresponding to the energy of a virtual defect associated with the unpaired moment $\vec{\mu}_0$, see Appendix~\ref{app:frustrated}. In this case, the invariance of the expression~\eqref{eq:exchange} with respect to the freedom of choice of $\vec{\mu}_0$ in the chain introduces antiperiodic boundary conditions on the N\'{e}el vector, $\vec{n}(0) \equiv -\vec{n}(L)$. This is also in agreement with theoretical~\cite{Castillo-Sepulveda17} and experimental~\cite{Cador04,Baker12} investigation of AFM rings with odd number of spins, where the M\"{o}bius magnetic structure appears as a special case of the AFM spin spiral in a geometrically periodic system. In this way, an equilibrium magnetic texture of $\vec{n}$ for chains with both signs of $J$ is determined by the variation of the total energy $E = E_\text{x}[\vec{n}] + E_\text{a}[\vec{n}]$, where the type of exchange bonds comes into the definition of the anisotropy only.

An interplay between exchange and anisotropy energies allows to introduce the effective magnetic length $\ell = \sqrt{A/|K|}$, which is a measure of spatial scales. In the following, the reduced coordinate $\xi = s/\ell$, chain length $X = L/\ell$, curvature $\varkappa = \kappa\ell$ and torsion $\sigma = \tau\ell$ are used. The aforementioned derivation of the continuum expressions for the magnetic energies is valid if $\ell \gg a$.

The exchange energy density in a curvilinear reference frame can be split into three terms, $E_\text{x}[\vec{n}] = |K|\ell \int (w_\text{x}^0 + w_\text{x}^\textsc{d} + w_\text{x}^\text{an}) d\xi$, {where} $w_\text{x}^0 = (n_\alpha')(n_\alpha')$ is the locally homogeneous part of exchange and prime means the derivative with respect to $\xi$, $w_\text{x}^\textsc{d} = \epsilon_{\alpha\beta\gamma}d_\alpha n_\beta n_\gamma'$ is the curvature-induced DMI and $w_\text{x}^\text{an} = \mathcal{F}_{\alpha\gamma}\mathcal{F}_{\beta\gamma}n_\alpha n_\beta$ is the curvature-induced anisotropy~\cite{Pylypovskyi20}. Here, the Einstein summation rule is used, Greek indices run over the local reference frame, $\alpha,\beta,\gamma=\overline{\mathrm{T,N,B}}$, $\epsilon_{\alpha\beta\gamma}$ is the totally antisymmetric Levi--Civita symbol, $\vec{d} = 2\sigma \vec{e}_\textsc{t} + 2\varkappa \vec{e}_\textsc{b}$ is the Dzyaloshinskii vector and $\mathcal{F}_{\alpha\beta}$ is the Frenet tensor with non-zero components $\mathcal{F}_\textsc{tn} = - \mathcal{F}_\textsc{tn} = \varkappa$ and $\mathcal{F}_\textsc{nb} = - \mathcal{F}_\textsc{bn} = \sigma$. The latter allows to define $w_\text{x}^\text{an} = k_{11}n_\textsc{t}^2 + k_{22}n_\textsc{n}^2 + k_{33}n_\textsc{b}^2 - k_{13}n_\textsc{t}n_\textsc{b}$ with $k_{11}(\xi) = \varkappa^2$, $k_{33} = \sigma^2$, $k_{22} = k_{11} + k_{33}$ and $k_{13}(\xi) = 2\varkappa\sigma$. The anisotropy $w_\text{x}^\text{an}$ is the source of the easy-axis anisotropy lying in the rectifying (TB) plane, in addition to the hard-axis anisotropy $E_\text{a}$ stemming from the dipolar interaction. A non-zero $\vec{d}$ can result in spin spiral states~\cite{Pylypovskyi20}. The distribution of $\vec{d}$ along the epicycle $\vec{\gamma}_{-3}^1$ is shown in Fig.~\ref{fig:first}(g). Because of the selection of the geometry, only the binormal component of $\vec{d}$ is coordinate-dependent. The strongest curvature-induced DMI appears at the knots, where $\vec{d}$ is almost aligned with the binormal direction, while in the remaining part of the closed curved
it follows the tangential direction.

\begin{figure*}[t]
	\includegraphics[width=\linewidth]{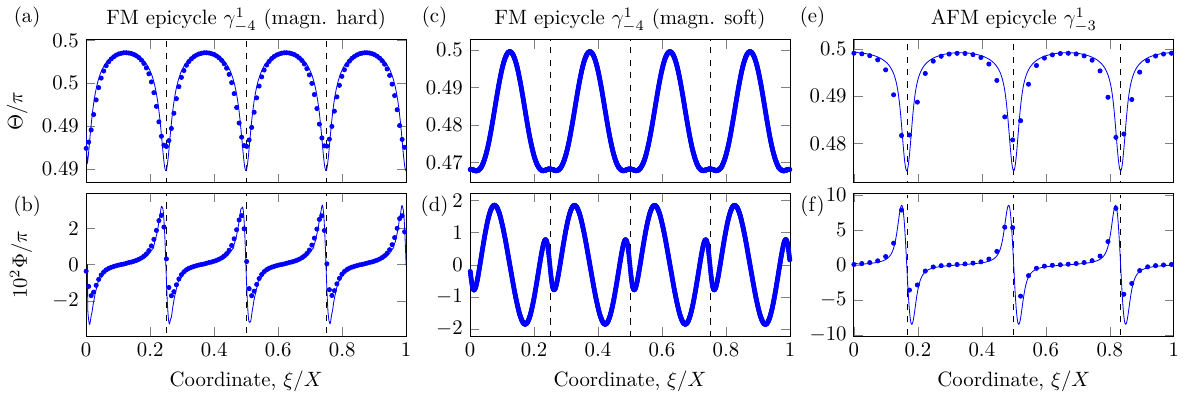}
	\caption{\textbf{Ground states of FM and AFM epicycles with the easy-axis anisotropy and $\sigma = 0.1$ (quasi-tangential state).} (a)~Polar and (b)~azimuthal angles of the order parameter $\vec{n}$ in a magnetically hard ($K_\text{dip} \ll K_\text{a}$) FM epicycle $\gamma_{-4}^1$. Symbols and solid lines correspond to simulations and Eq.~\eqref{eq:ea-solution}, respectively. (c,d) The same for a magnetically soft FM epicycle $\gamma_{-4}^1$ ($K_\text{a} = 0$). Only simulation results are shown. The magnetic state is perturbed by the long-range part of the dipolar interaction. (e,f) The same for an AFM epicycle $\gamma_{-3}^1$ with the dominating single-ion anisotropy. In panels (a,b,e,f) each 10-th symbol in simulations is shown. In all panels, the black dashed lines show positions, where the curvature is maximal.}
	\label{fig:ea-epicycles}
\end{figure*}

\section{Magnetic states}
\label{sec:magnetic-states}

A static state within the nonlinear $\sigma$-model described in Sec.~\ref{sec:sigma-model} is not sensitive to the sign of $J$. Still, it strongly depends on the (anti)periodicity of the boundary conditions. Therefore, first we consider two cases of the easy- and hard-axis anisotropy in FM and AFM epicycles and then analyze the frustrated epicycles with both cases of the anisotropy coefficient $K$. In this section, we focus on the case of small torsion, $|\sigma| \ll 1$, which corresponds to values of $\mathscr{R}$ large enough for anisotropic effects to dominate over exchange ones. For large torsions, spin chains can experience the helimagnetic phase transition~\cite{Sheka15a,Pylypovskyi20}, which is discussed in Appendix~\ref{app:simulations}.

\subsection{FM and AFM epicycles with the easy axis anisotropy}
\label{sec:fm-afm-ea}

In this section, we focus on the FM and AFM epicycles $\gamma_p^1$ with the easy-axis anisotropy ($K > 0$). We start from the case of a magnetically hard system, with the contribution of anisotropy being much stronger than the dipolar interaction, i.e., $K_\text{a} \gg K_\text{dip}$. It is convenient to introduce the local spherical parameterization of the order parameter as
\begin{equation}\label{eq:ea-magnetic-angles}
	\vec{n} = \sin\Theta_\text{ea}(\cos\Phi_\text{ea} \vec{e}_\textsc{t} + \sin\Phi_\text{ea} \vec{e}_\textsc{n}) + \cos\Theta_\text{ea} \vec{e}_\textsc{b}
\end{equation}
with $\Theta_\text{ea}(\xi)$ and $\Phi_\text{ea}(\xi)$ being magnetic polar and azimuthal angles following the periodic boundary conditions. Here and in the following, the magnetic polar angles are defined within $[0,\pi]$ range and the magnetic azimuthal angles can take arbitrary values. For $|\sigma|\ll1$, the ground state reads $\Theta_\text{ea} = \pi/2 + \vartheta_\text{ea}(\xi)$ and $\Phi_\text{ea} = 0.5(1\mp 1)\pi + \varphi_\text{ea}(\xi)$, where $\vartheta_\text{ea},\varphi_\text{ea} \ll1$ satisfy equations of the nonlinear pendulum with an external drive and parametric pumping
\begin{subequations}\label{eq:fm-equations}
	\begin{equation}\label{eq:fm-vartheta}
		\vartheta_\text{ea}'' - [1 + k_{11}(\xi)] \vartheta_\text{ea}  = \pm k_{13}(\xi) \equiv \pm \sigma \varkappa(\xi), 
	\end{equation}
	\begin{equation}\label{eq:fm-varphi}
		\varphi_\text{ea}'' - \varphi_\text{ea} = - \dfrac{d_\textsc{b}'(\xi)}{2} \equiv -\varkappa'(\xi).
	\end{equation}
\end{subequations}
These angles also satisfy the periodic boundary conditions $\vartheta_\text{ea}(0) = \vartheta_\text{ea}(X)$ and $\varphi_\text{ea}(0) = \varphi_\text{ea}(X)$. Neglecting parametric pumping determined by $k_{11}(\xi)$ in Eq.~\eqref{eq:fm-vartheta} which also implies that $\varkappa(\xi) \ll 1$, the system~\eqref{eq:fm-equations} can be solved asymptotically. The respective polar and azimuthal angles of the order parameter read
\begin{equation}\label{eq:ea-solution}
	\begin{split}
		\Theta_\text{ea}(\xi) & = \dfrac{\pi}{2} \mp \sigma \varkappa(\xi) + \mathcal{O}(\varkappa^2), \\
		\Phi_\text{ea}(\xi) &= \dfrac{1\mp 1}{2}\pi + \varkappa'(\xi) + \mathcal{O}\left(\dfrac{|\varkappa'''|}{|\varkappa'|}\right).
	\end{split}
\end{equation}
We refer to this state as the quasi-tangential state. Naturally, the main deviation from the purely tangential state occurs in the vicinity of the knots of the epicycle, corresponding to the maximum of curvature. The comparison of the solution~\eqref{eq:ea-solution} with simulations is shown in Fig.~\ref{fig:ea-epicycles}. As it follows from numerical analysis, the solution~\eqref{eq:ea-solution} matches the simulation data with a good accuracy, even for $\varkappa(\xi) \lesssim 1$. For both FM and AFM epicycles with $\sigma = 0.1$ [Fig.~\ref{fig:ea-epicycles}(a,b,e,f)], there is a small difference between analytics and simulations near the maxima of curvature, where $\varkappa_\text{max} \approx 0.48$ for the FM epicycle $\gamma_{-4}^1$  and $\varkappa_\text{max} \approx 1$ for the AFM epicycle $\gamma_{-3}^1$. 

\begin{figure*}[t]
	\includegraphics[width=\linewidth]{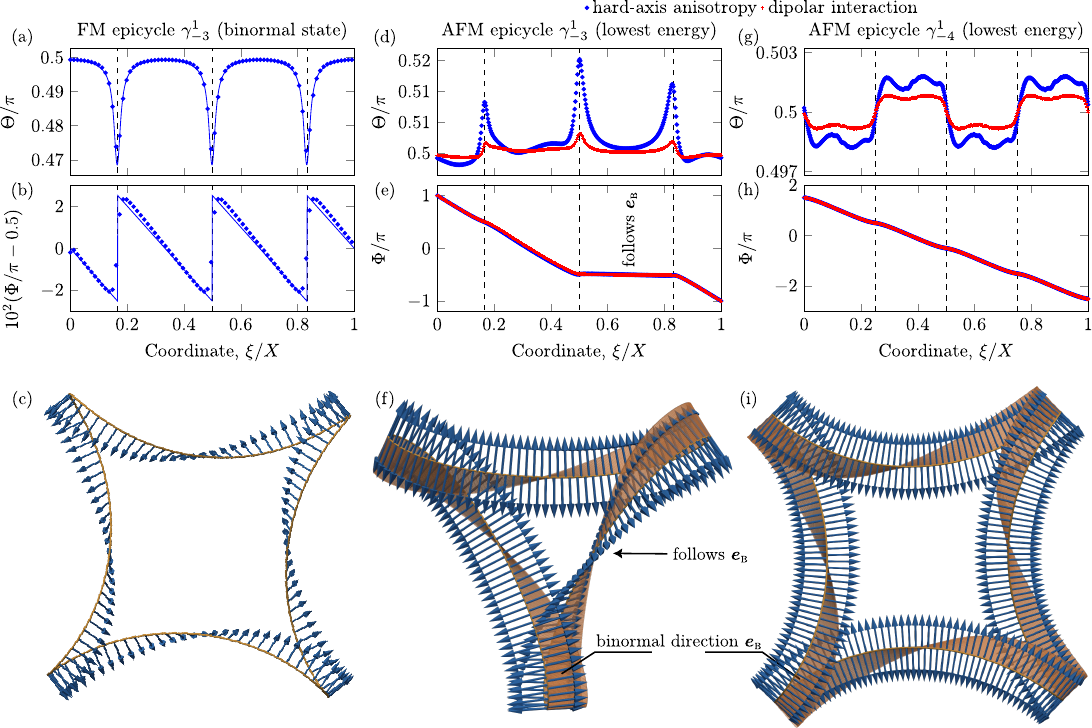}
	\caption{\textbf{Magnetic states of FM and AFM epicycles with the hard-axis anisotropy and $\sigma = 0.1$.} (a)~Polar and (b)~azimuthal angles of the order parameter $\vec{n}$ in the FM epicycle $\gamma_{-3}^1$. Symbols (each 10-th simulated datapoint is shown) and solid lines correspond to simulations and Eq.~\eqref{eq:ha-solution-binorm}, respectively. The black dashed lines show positions, where the curvature is maximal. (c)~The quasi-binormal magnetic state of the FM epicycle $\gamma_{-4}^1$ in simulations, c.f.~Fig.~\ref{fig:first}(g). The blue arrows correspond to the direction of the each 3-rd local magnetic moment. (d--f)~and (g--i) show the same for the $2\pi$-twist and $4\pi$-twist states (lowest-energy states) of the AFM epicycles $\gamma_{-3}^1$ [c.f.~Fig.~\ref{fig:first}(a)] and $\gamma_{-4}^1$. The blue symbols and red crosses correspond to the cases of the hard axis anisotropy and dipolar interaction, respectively. In panels (f,i) blue arrows show equidistant magnetic moments from the opposite sublattices (each 3-rd site is shown). A semi-transparent yellow surface indicates the direction of $\vec{e}_\textsc{b}$, which coincides with the magnetic moment's axes only within certain segments of the epicycle indicated in~(f).
	}
	\label{fig:ha-epicycles}
\end{figure*}

The ground state of the magnetically soft epicycle ($K_\text{a} = 0$) is also close to the tangential state, although the magnetic state cannot be described by Eqns.~\eqref{eq:ea-solution}, see Fig.~\ref{fig:ea-epicycles}(c,d). Unlike the aforementioned case of hard magnetic material, both spherical angles deviate from the tangential
direction in a spread region around the maxima of curvature. In particular, the spatial distribution of the azimuthal angle has wave vectors corresponding to the wavelengths $X/|p|$ and $2X/|p|$. This behavior originates from the long-range part of the dipolar interaction, which leads to the self-interaction of the neighboring parts of the epicycle similarly to the experimental observations in planar~\cite{OBrien11} and 3D ferromagnetic geometries~\cite{Donnelly22}. We note that the quasi-tangential ground state is an analogy of the vortex state in planar rings with easy-axis anisotropy~\cite{Sheka15}.

\subsection{FM and AFM epicycles with the hard axis anisotropy}
\label{sec:fm-afm-ha}

In this section, we focus on the FM and AFM epicycles $\gamma_p^1$ with the hard-axis anisotropy ($K < 0$). In comparison with the easy-axis magnets (Sec.~\ref{sec:fm-afm-ea}), a possibility to pin an inhomogeneous magnetic texture at the epicycle knots supports a variety of metastable states and complicates the determination of the ground state. Therefore, in the following we describe the lowest-energy states numerically found. The discussion of a possibility to have these states as the ground state is given in Sec.~\ref{sec:discussion}. 

A convenient angular parameterization for the order parameter in the system with the hard tangential axis of anisotropy is
\begin{equation}\label{eq:ha-magnetic-angles}
	\vec{n} = \cos\Theta_\text{ha} \vec{e}_\textsc{t} + \sin\Theta_\text{ha}(\cos\Phi_\text{ha} \vec{e}_\textsc{n} + \sin\Phi_\text{ha} \vec{e}_\textsc{b})
\end{equation}
with the polar and azimuthal angles $\Theta_\text{ha}$ and $\Phi_\text{ha}$, respectively, which follow the periodic boundary conditions. As it is known for AFM rings with the even number of spins and AFM helices~\cite{Pylypovskyi20}, their ground state is close to the binormal one with $\Theta_\text{ha} = \Phi_\text{ha} = \pi/2$. This happens due to the appearance of a weak easy-axis anisotropy of the exchange origin with the coefficient $k_{11}(\xi)$. The corresponding static state on the background of the binormal state, $\Theta_\text{ha} = \pi/2 + \vartheta_\text{ha}(\xi)$ and 
$\Phi_\text{ha} = \pm \pi/2 + \varphi_\text{ha}(\xi)$
, where $\vartheta_\text{ha},\varphi_\text{ha} \ll1$, satisfies equations
\begin{subequations}\label{eq:afm-equations-binorm}
	\begin{equation}\label{eq:afm-vartheta-binorm}
		\vartheta_\text{ha}'' -  [1 + k_{11}(\xi)] \vartheta_\text{ha} = \pm \left( k_{13}(\xi) - \dfrac{d_\textsc{b}'(\xi)}{2} + \dfrac{d_\textsc{b}(\xi)\varphi_\text{ha}'}{2}\right), 
	\end{equation}
	\begin{equation}\label{eq:afm-varphi-binorm}
		\varphi_\text{ha}'' - k_{11}(\xi) \varphi_\text{ha} = \mp  \dfrac{[\vartheta_\text{ha}d_\textsc{b}(\xi)]'}{2}
	\end{equation}
\end{subequations}
with the periodic boundary conditions $\vartheta_\text{ha}(0) = \vartheta_\text{ha}(X)$ and $\varphi_\text{ha}(0) = \varphi_\text{ha}(X)$. Similarly to the case of FM epicycles, the polar angle in the binormal state can be found asymptotically. The azimuthal angle is a solution of the equation with strongly nonlinear coefficients and can be approximated by a linear function. Together they read
\begin{equation}\label{eq:ha-solution-binorm}
	\begin{split}
		\Theta_\text{ha}(\xi) & = \dfrac{\pi}{2} \mp \sigma \varkappa(\xi) + \mathcal{O}(\varkappa^2), \\
		\Phi_\text{ha}(\xi) & \approx \pm \dfrac{\pi}{2} + \dfrac{\sigma}{2}\left[\dfrac{\pi}{2} - \left( p\pi \dfrac{\xi}{X}- \dfrac{\pi}{2} \right) \mathrm{mod}\,\pi \right].
	\end{split}
\end{equation}
This expression fits the spatial distribution of the order parameter for FM and AFM epicycles, see Fig.~\ref{fig:ha-epicycles}(a--c). We refer to the state~\eqref{eq:ha-solution-binorm} as the quasi-binormal state. 

\begin{figure*}[t]
	\includegraphics[width=\linewidth]{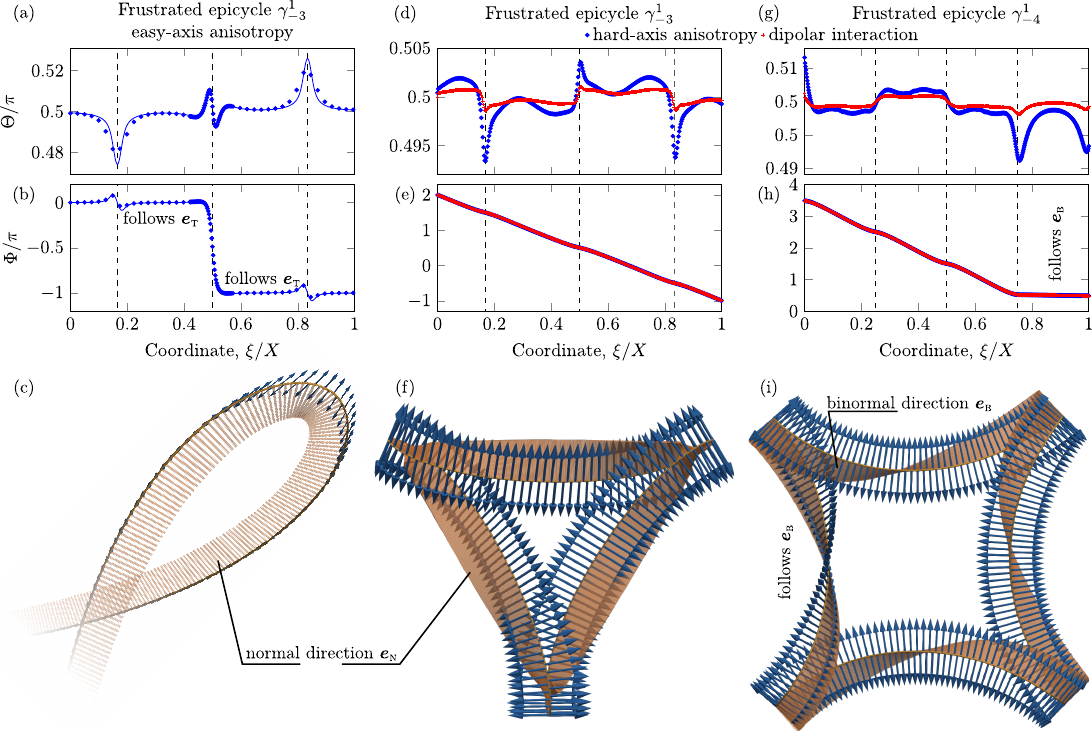}
	\caption{\textbf{Ground states of frustrated epicycles with the easy-axis anisotropy and $\sigma = 0.1$.} (a)~Polar and (b)~azimuthal angles of the order parameter $\vec{n}$ in a frustrated epicycle $\gamma_{-3}^1$ with the easy-axis anisotropy. Symbols and solid lines (drawn outside the domain wall region) correspond to simulations and Eq.~\eqref{eq:ea-solution}, respectively. Outside the domain wall, each 10-th simulated dataport is shown. (c) Zoom into the knot hosting the domain wall in simulations. Blue arrows show individual magnetic moments. Yellow cylinders indicate the normal direction $\vec{e}_\textsc{n}$. Panels (d--f) and (g--i) the same for the frustrated epicycles $\gamma_{-3}^1$ and $\gamma_{-4}^1$ with the hard-axis anisotropy (blue symbols) and dipolar interaction (red crosses), respectively ($3\pi$-twist state). In panels (f) and (g), a semitransparent yellow surface indicates the direction of $\vec{e}_\textsc{n}$ and $\vec{e}_\textsc{b}$, respectively. Each 3-rd magnetic moment is shown.}
	\label{fig:fru-epicycles}
\end{figure*}

Surprisingly, the state~\eqref{eq:ha-solution-binorm} is not the ground state for these epicycles. The lowest energy state of the epicycle $\gamma_{-3}^1$ consists of two spatial regions with different behavior of the order parameter [Fig.~\ref{fig:ha-epicycles}(e--f)]. There is a segment of the epicycle in which the order parameter follows the binormal direction, which looks as a twist drawn by $\vec{n}$ in space [Fig.~\ref{fig:first}(a) and Fig.~\ref{fig:ha-epicycles}(f)]. In the remaining part of the geometry, the order parameter slowly rotates by $\pi$ radians around $\vec{e}_\textsc{t}$ on each of the segments, which looks as the almost uniform distribution of $\vec{n}$ in the laboratory reference frame ($2\pi$-twist state). 

The AFM epicycle $\gamma_{-4}^1$ [Fig.~\ref{fig:ha-epicycles}(g--i)] in the lowest-energy state has a $\pi$ twist of the order parameter $\vec{n}$ within each of its segments. The respective $4\pi$-twist state is shown in Fig.~\ref{fig:ha-epicycles}(g--i). This geometry also supports $2\pi$-twist states with two segments following the quasi-binormal state. Among $2\pi$-twist states, the most symmetric case with quasi-binormal distribution at the oppositely placed segments has lowest energy. If the quasi-binormal distribution is obtained in two sequential states, the energy is slightly larger because of the modification of the magnetic texture at the knots.

We stress that the discussed states of the epicycles with easy and hard axis of anisotropy are the same in terms of the order parameter $\vec{n}$ for both FM and AFM cases considered in these sections. For all of them the excited states include domain walls pinned at the epicycle knots. 

In case of the absence of the single-ion anisotropy in spin chains with the AFM exchange bonds, the dipolar interaction leads to an effective hard-axis anisotropy. We found that in this case, the azimuthal angle $\Phi$ agrees well with the results of simulations with the hard-axis anisotropy only [Fig.~\ref{fig:ha-epicycles}(d,e,g,h), red crosses]. The difference is in the amplitude of the change of the polar angle $\Theta$, which is smaller for the case of dipolar interaction. The simulations are done for the nominally same magnetic lengths in both cases. Thus, we expect that the origin of the difference between $\Theta(\xi)$ in these simulations arises due to the partial uncompensation of the antiferromagnetically ordered spins near the epicycle knots similarly to FM epicycles [c.f. Fig.~\ref{fig:ea-epicycles}(d)].

\subsection{Magnetic states in frustrated epicycles}

In this section we consider epicycles with the AFM nearest-neigbor exchange and odd number of spins. Their magnetic states follow the same equations for the angular variables introduced in~\eqref{eq:ea-magnetic-angles} and~\eqref{eq:ha-magnetic-angles} for the easy- and hard-axis anisotropies, respectively. The difference is that here the antiperiodic boundary conditions are applied for $\vec{n}$, i.e., $\Theta_\text{ea,ha}(0) = \pi - \Theta_\text{ea,ha}(X)$ and $\Phi_\text{ea,ha}(0) = \Phi_\text{ea,ha}(X) + \pi + 2\pi w$ with $w \in \mathbb{Z}$. For the case of planar rings, the ground state corresponds to the so-called M\"{o}bius state with the order parameter making one twist around the circle~\cite{Castillo-Sepulveda17}.

For frustrated epicycles with the easy-axis anisotropy, the antiperiodic boundary conditions impose the appearance of a domain wall in the ground state. The domain wall is located at one of the knots because of the strong curvature-induced DMI at this location. Far from the domain wall, the magnetic state can be approximately described by Eqns.~\eqref{eq:ea-solution}, see Fig.~\ref{fig:fru-epicycles}(a--c). 

Hard-axis frustrated epicycles with even and odd number of knots have different lowest-energy states. In the lowest-energy state, the magnetic texture of each of three segments of the $\gamma_{-3}^1$ epicycle is almost uniform in the laboratory reference frame, which corresponds to the rotation of $\vec{n}$ by $\pi$ radians around the tangential direction ($3\pi$-twist state) [Fig.~\ref{fig:fru-epicycles}(d--f)]. The epicycle with four knots  in the lowest-energy state has $3\pi$-twist over three segments, and one segment in the quasi-binormal state  [Fig.~\ref{fig:fru-epicycles}(g--i)]. We note that in case of the change of the hard-axis anisotropy to the dipolar interaction in simulations, the observed changes in the magnetic state are the same as for AFM epicycles, c.f. Fig.~\ref{fig:ha-epicycles}(d,e,g,h) and~\ref{fig:fru-epicycles}(d,e,g,h).

\section{Ferrotorodial ordering}
\label{sec:ferrotoroidal}

\begin{figure*}
	\includegraphics[width=\linewidth]{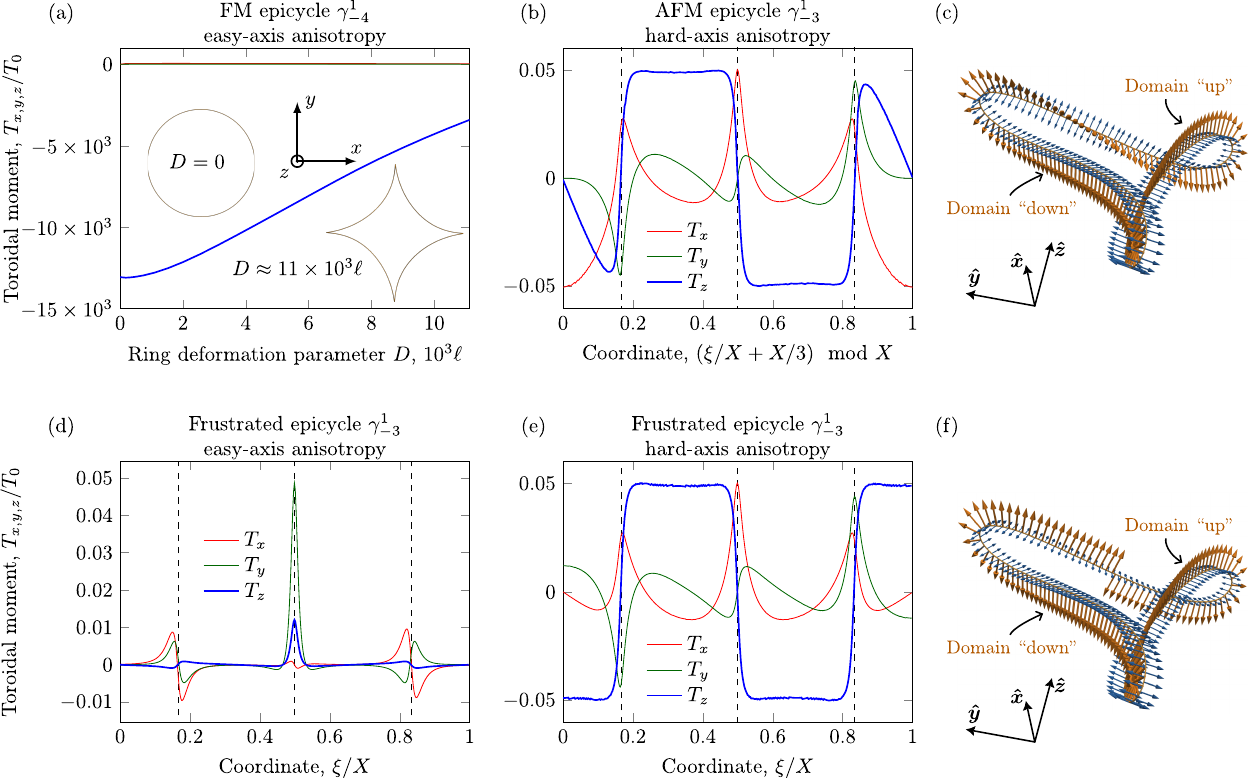}
	\caption{\textbf{Ferrotoroidal ordering in magnetic epicycles.} (a)~The total toroidal moment of a ferromagnetic spin chain deformed from a ring ($D = 0$) to the epicycle $\gamma_{-4}^1$ [$D \approx 11\times 10^3$, the same as one shown in Fig.~\ref{fig:ea-epicycles}(a,b)]. Insets show the axial views of the respective geometries. The coordinate system $(x,y,z)$ is shown as well. (b)~Local toroidal moment in the AFM epicycle $\gamma_{-3}^1$ with the hard-axis anisotropy and (d) its 3D view with the blue and orange arrows showing the magnetic and toroidal moments, respectively (each 3-rd site is shown). The corresponding state is shown in Fig.~\ref{fig:ha-epicycles}(d--f); note the shift by $X/3$ in coordinate. (d)~The same for a frustrated epicycle~$\gamma_{-3}^1$ with the easy-axis anisotropy. The corresponding state is shown in Fig.~\ref{fig:fru-epicycles}(a--c). (e,f) The same for a frustrated epicycle~$\gamma_{-3}^1$ with the hard-axis anisotropy. The corresponding state is shown in Fig.~\ref{fig:fru-epicycles}(d--f). 
	}
	\label{fig:toroidal}
\end{figure*}

The here-studied closed-loop spin chains lack both space-inversion and time-reversal symmetries. This makes them potential host for magnetoelectric multipoles defined by~\cite{Ortix23} ${\mathcal M}_{ij}= \int {\bf r}_i \boldsymbol{\mu}_j({\bf r}) d^3 {\bf r}$ that can be decomposed into three irreducible tensors: the pseudoscalar $A= \textrm{tr} {\mathcal M}_{ij} /3$ defining the magnetoelectric monopole~\cite{Spaldin13}; the toroidal moment related to the antisymmetric part ${\bf T}=\epsilon_{ijk} M_{jk}/2$; and the traceless symmetric tensor describing the quadrupole magnetic moment. These three components of the ${\mathcal M}_{ij}$ tensor are of paramount importance to the linear magnetoelectric response $\|\alpha_{ij}\|$, $i,j=x,y,z$. 
Before computing the magnetoelectric multipoles in our closed-loop spin chains, we recall that one of the main complications in their definition is their origin dependence for uncompensated magnetic textures. In the following, we will take into account only the contribution from the \textit{compensated} part of the spin texture that is instead origin independent.

First, we use a Landau field theory that incorporates ferroelectric polarization $\vec P$ and magnetization $\vec M$. Since the epicycles $\gamma_{-3}^1$ and $\gamma_{-4}^1$ are characterized by the presence of $3_z$ or $4_z$ axes of rotation, respectively, the free energy terms for the variable $\vec{O}$ with $\vec{O} = \vec{M}$ or $\vec{P}$ are the same. In particular, the free energy allows terms $O^2$ and $O_z^2$ of the second order, terms $M^4$, $P^4$, $M^2P^2$, $(\vec{M}\cdot \vec{P})^2$ of the fourth order. The magnetoelectric coupling term reads
\begin{equation}\label{eq:free-energy-me}
	\begin{aligned}
		F_\textsc{me} & = \alpha_{xx} (M_xP_x + M_y P_y) + \alpha_{zz} M_zP_z \\
		& + \alpha_{xy} (M_xP_y - M_y P_x)
	\end{aligned}
\end{equation}
with $\vec{\hat{z}}$ axis along the symmetry axis of the structure. The first two terms in Eq.~\eqref{eq:free-energy-me}, as the diagonal part of the magnetoelectric coupling, can be decomposed in the magnetoelectric monopolization $a$ and the $q_{x^2},q_{y^2}$ and $q_{z^2}$ components of the quadrupolar moment with the additional constraint $q_{x^2} \equiv q_{y^2}$ and the traceless condition $q_{x^2}+q_{y^2}+q_{z^2} \equiv 0$. The last term instead indicates the magnetoelectric coupling corresponding to the toroidal moment $\vec{T}$ directed along $\vec{\hat{z}}$. 
The other components of the toroidal moment as well as the quadrupolar moment components $q_{ij}$ with $i \neq j$ are not allowed by symmetry. 

We proceed with the calculation of the change in the total toroidal moment in a FM system for the smooth change of the geometry of epicycle into a ring, see Fig.~\ref{fig:toroidal}(a). The geometry transformation is calculated by mimicking the mechanical forces acting in the plane perpendicular to the epicycle's axis. To describe these changes, we characterize each geometry by the deformation parameter $D = \sum_{i=0}^{N-1} |\vec{r}_i - \vec{r}_i^\text{circ}|$, where $\vec{r}_i^\text{circ}$ is the radius-vector of the $i$-th spin at the ring. In this way, the quasi-tangential state~\eqref{eq:ea-solution} is smoothly translated into a vortex state on the ring~\cite{Sheka15}. Selecting the origin for the radius-vector of the $i$-th moment $\vec{r}_i$ in the center of mass of the sample, we obtain the origin-independent value of the toroidal moment relying by accounting for the compensated part of the magnetization only. Fig.~\ref{fig:toroidal}(a) shows that $T_z(D)$ smoothly changes with the geometry transformation in agreement with the symmetry of the free energy~\eqref{eq:free-energy-me}. The numerical values of the in-plane components of the toroidal moment, as well as of the monopole moment and components of the quadrupole moments, are of the same order and about 200 times smaller than $T_z$. We correlate their finite values with the discrete structure of the spin chain.

The fact that only the $z$ component of the toroidal moment is non-vanishing implies that the spin texture realizes a purely toroidal configuration, similarly to Bloch skyrmions~\cite{Bhowal22}. 
We note that for a vortex state on a ring, this purely toroidal state is symmetry enforced. The vertical mirror symmetries imply that the magnetoelectric monopole, as well as the quadrupole moment components $q_{x^2,y^2,z^2}$ must vanish. In addition, the $T_{x,y}$ components are zero due to the planar spin configuration. In the epicycles instead these symmetries are not present. Therefore, the purely toroidal configuration is spontanelously chosen by the spin texture. We note that the case of a hard-axis FM epicycle is more complex and the corresponding magnetic state can support finite monopole or quadrupolar contributions in addition to the toroidal moment.

In the following, we discuss magnetoelectric monopoles in AFM and frustrated epicyles. As magnetic monopoles of the two magnetic ``sublattices" are equal and opposite, the \textit{total} value of the magnetoelectric monopoles is negligibly small. Still, the \textit{local} values of the magnetoelectric monopoles from each pair of magnetic moments can be different to zero. For instance, the local value of $\vec{T}(\xi)$ can be calculated for the each pair $\vec{\mu}_{2i-1,2i}$ on which $\vec{n}_i$  is introduced with respect to its center: 
\begin{equation}
	\begin{aligned}
		\vec{t}_i & = \dfrac{1}{2\ell}[(\vec{r}_{2i-1}-\vec{r}_i^c)\times \vec{\mu}_{2i-1} + (\vec{r}_{2i}-\vec{r}_i^c)\times \vec{\mu}_{2i}] \\
		 & - \dfrac{1}{2\ell}\vec{r}_i^c \times (\vec{\mu}_{2i-1}+\vec{\mu}_{2i})
	\end{aligned}
\end{equation}
with $\vec{r}_i^c = 0.5(\vec{r}_{2i-1}+\vec{r}_{2i})$. AFM epicycles with the odd number of knots show the presence of well-defined toroidal domains in the laboratory reference frame oriented along the axis of symmetry located between the epicycle's knots, see the blue line in Fig.~\ref{fig:toroidal}(b) and Fig.~\ref{fig:toroidal}(c) for the epicycle $\gamma_{-3}^1$. Here, the opposite toroidal domains correspond to the almost linearly changing $\Phi$ in Fig.~\ref{fig:ha-epicycles}(e). The same picture with three sequential toroidal domains is observed for the frustrated epicycle $\gamma_{-4}^1$. 

An easy-axis frustrated epicycle shows a pronounced change of $\vec{T}$ only at knots with the maximum value reached at the domain wall [Fig.~\ref{fig:toroidal}(d)]. This finding is in agreement with the expectation of the finite magnetization at the inhomogeneous distribution of the N\'{e}el vector in spin chains~\cite{Pylypovskyi21e}. The symmetry of $\vec{T}(\xi)$ in a hard-axis frustrated epicycle $\gamma_{-3}^1$ [Fig.~\ref{fig:toroidal}(e,f)] is different from the AFM epicycle with three knots [Fig.~\ref{fig:toroidal}(b)] due to antiperiodic boundary conditions.
These distinct features are expected to affect the local ferroelectric polarization that can be induced by externally applying a magnetic field via the trilinear coupling ${\vec T} \times {\vec P} \cdot {\vec M} $. For instance, for AFM epicycles with three knots, we expect ferroelectric domains with equal and opposite polarization along the $\vec{\hat{x}}$ direction if a constant magnetic field along the $\vec{\hat{y}}$ direction is applied. 

\section{Discussion}
\label{sec:discussion}

Here, we described static magnetic states in curvilinear closed spin chains with constant torsion $\sigma$. We focused on cases of the easy- and hard-axis anisotropies in spin chains with FM nearest-neighbor exchange as well as chains with AFM exchange containing odd or even number of spins. The particular case of space curves with constant torsion discussed in this work complements the results known for planar rings and helices, both with $\varkappa = \text{const}$ and $\sigma = \text{const}$. The case of closed curves allows us to focus on the geometries topologically related to planar rings and track the change of their magnetic textures with the change of geometry. Nevertheless, unlike planar rings, the Dzyaloshinskii vector $\vec{d}$ in curves of constant torsion always has a non-vanishing tangential component, which may compete with other curvature-induced and intrinsic interactions.

For easy-axis FM and AFM epicycles, the ground state is the quasi-tangential state with the order parameter $\vec{n}$ directed almost along the tangential direction with a small deviation near the knots, Fig.~\ref{fig:ea-epicycles}. The effects of a finite torsion are pronounced in hard-axis samples, where the geometry-driven Dzyaloshinskii vector $\vec{d}$ is almost perpendicular to the order parameter $\vec{n}$ on a major part of the spin chain. Hence, it can lead to a spin spiral and topologically non-trivial states when taking into account the boundary conditions for $\vec{n}$ in closed chains. The lowest-energy state for hard-axis AFM epicycles with even number of spins and FM epicycles is more complex (Fig.~\ref{fig:ha-epicycles}) and can be understood by comparing with frustated epicycles which have AFM exchange and odd number of spins (Fig.~\ref{fig:fru-epicycles}). The local state in each geometric segment of the epicycle is determined by the following aspects: (i)~The condition of a constant torsion $\sigma$ implies a constant tangential component of the curvature-induced Dzyaloshinskii vector $\vec{d}$. (ii)~The curvature changes in a wide range of values from $\varkappa \ll \sigma$ at the center of the segment to $\varkappa \gtrsim \sigma$ at its ends, see Fig.~\ref{fig:first}(b). Thus, unlike the geometry-driven anisotropy, the geometry-driven DMI is never vanishing and can lead to a spin spiral state.

The local ground state for the case of constant or very slowly varying $\sigma$ and $\varkappa$ is determined by their balance: for $\sigma \lesssim \varkappa$, the state in the local reference frame is uniform (twisted in the laboratory reference frame). Otherwise, the spin chain is in a helicoidal state, which is almost uniform in the laboratory reference frame~\cite{Pylypovskyi20}. Thus, in the central part of the epicycle's segment, the curvature-induced DMI dominates over the curvature-induced anisotropy, thus favoring rotations of $\vec{n}$ around $\vec{d} \approx \vec{e}_\textsc{t}$. At the segment's ends, the physical picture is opposite: the curvature-induced anisotropy $w_\text{x}^\text{an}$ determined by $\varkappa$ cannot be considered being small and induces an easy axis along $\vec{e}_\textsc{b}$. Together with a strong curvature-induced DMI $\vec{d}\approx 2\varkappa \vec{e}_\textsc{b}$, the curvature-induced anisotropy produces a pinning potential for inhomogeneous magnetic textures.

Therefore, the segments can support the following magnetic states: (i) binormal state if the energy contribution from knots dominates; (ii) helicoidal if the energy contribution from the central part dominates; (iii) mixed state resembling freatures of the binomial and helicoidal states. Although the size of the epicycle's segment and the helicoid period $\sim 1/\sigma$, the only possible rotation of $\vec{n}$ corresponding to the minimal energy corresponds to the twist by $\pi$ radians. The total change of the magnetic azimuthal angle $\Phi$ over the whole geometry in AFM and FM epicycles should be a multiple of $2\pi$ to match the periodic boundary conditions. The segments accommodating the change of $\Phi$ by $\pi$ are almost magnetically uniform in the laboratory reference frame [top and left segments of the $\gamma_{-3}^1$ epicycle in Fig.~\ref{fig:ha-epicycles}(f), and all segments of the $\gamma_{-4}^1$ epicycle in Fig.~\ref{fig:ha-epicycles}(i)]. The energy contribution from knots is important for small-size epicycles (large $\sigma$, which leads to a much larger $\varkappa_{\max}$), see Appendix~\ref{app:simulations}. In this case, the lowest-energy state corresponds to an almost uniform texture in the laboratory reference frame with $\vec{n}$ directed along the axis of symmetry. We note that this state corresponds to a helicoidal state in the local reference frame. 

Frustrated epicycles maintain the total phase equal $(2w+1)\pi$ with $w \in \mathbb{N}$, which allows the texture to fit the antiperiodic boundary conditions. Since the $\pi$-twist state (with two segments being in a quasi-binormal state) of a frustrated epicycle $\gamma_{-3}^1$ has higher energy than $3\pi$-twist state for the geometries under consideration, the helicoidal texture within the segment is energetically more preferable than the binormal state. This is an indication that the influence of the curvature-driven DMI dominates over the anisotropy in the magnetic energy. Based on the symmetry considerations, the described lowest-energy states for FM and AFM epicycles can be considered as the ground states for the geometries $\gamma_{-3}^1$ and $\gamma_{-4}^1$. This conclusion can be extrapolated to epicycles with $|p| > 4$ if their segment length is much larger than $\ell$.

For easy-axis systems, only the case of magnetically soft FM epicycles is special because of the self-interaction between the epicycle's segments via the dipolar interaction. The hard-axis systems reveal more specific geometry-dependent behavior. Here, there are two origins of the geometric frustration. The first one is the odd number of spins relevant for AFM-coupled spin chains [Fig.~\ref{fig:fru-epicycles}(d--i)], which enforces the antiperiodic boundary condition for $\vec{n}$. The second origin of frustration is introduced by the epicycle knots separating regions, which may have different magnetic textures. In this case, it is possible to track how the specific magnetic states can emerge either due to the even (odd) number of spins, or due to odd (even) number of knots in the epicycle. In particular, the $2\pi$-state, which contains a segment in a quasi-binormal state, in an AFM epicycle with the odd number of knots $\gamma_{-3}^1$ [Fig.~\ref{fig:ha-epicycles}(d--f)] resembles a frustrated epicycle with the even number of knots $\gamma_{-4}^1$ [Fig.~\ref{fig:fru-epicycles}(g--i)] and vice versa [c.f. Figs.~\ref{fig:ha-epicycles}(g--i) and~\ref{fig:fru-epicycles}(d--f)]. As a rule of thumb, the state of a hard-axis spin chain is magnetically frustrated if the sum of knots and spins $Q = (|p| + N) \mod 2$ is nonzero. 

A specific feature of the geometries with (anti)periodic boundary conditions is the possibility to develop a finite toroidal moment $\vec{T}$, which provides a bridge between magnetic and electric degrees of freedom~\cite{Murray22}. In this respect, easy-axis ferromagnetic spin chains can be characterized by a macrotoroidal moment depending on their shape [Fig.~\ref{fig:toroidal}(a)]. Hard-axis epicycles split into a multi-domain state by $\vec{T}$ with a sizeable axial component $T_z$. The aforementioned interplay between the frustration of the magnetic state produced either by the odd number of spins or the odd number of knots is also reflected in the formation of toroidal domains: a series of oppositely oriented domains appears for $Q = 1$.

To summarize, the presented analysis shows that the geometric frustration in magnetic systems can appear due to specific distributions of the curvature and torsion in a spin chain. This is a new route towards magnetic frustration, which is complementary to the established possibilities due to the local environment of each spin with the negative Toulouse frustration function~\cite{Toulouse77,Toulouse81} or antiperiodic boundary conditions for the magnetic order parameter~\cite{Castillo-Sepulveda17}. A fingerprint of the competition between different origins of frustations can be observed in the inhomogeneity of the magnetic order parameter $\vec{n}$ and local ferrotoroidic response.

We anticipate that our findings can be used to interpret magnetic responses in molecular magnets and complex 3D magnetic architectures fabricated, e.g., by means of glancing angle deposition (GLAD)~\cite{Phatak14,Gibbs14,Askey24} or focused electron beam induced deposition (FEBID)~\cite{Phatak20,Donnelly22}. Furthemore, this theory analysis can be extended to other geometries with specific symmetry properties like Salkowski curves~\cite{Monterde09,Monterde24} and guide experimental efforts on the design of the sample shapes to realize ferrotoroidic behavior. From the fundamental point of view, we expect that complementary effects may be observed in 3D magnetic nanostructures with constant curvature, where the Dzyaloshinskii vector $\vec{d}$ is characterized by a non-vanishing binormal component.

\section{Acknowledgments}

We thank Dr. Oleksii M. Volkov (HZDR) and Prof.~Denis~D.~Sheka (Taras Shevchenko National Univeristy of Kyiv) for fruitful discussions on curvilinear spin chains. This work is supported in part via German Research Foundation (grants MA5144/22-1, MA5144/24-1) and ERC grant 3DmultiFerro (Project number: 101141331). We acknowledge partial support by the Italian Ministry of Foreign Affairs and International Cooperation, grant PGR12351 ``ULTRAQMAT". Numerical calculations have been performed using the Hemera high-performance cluster at the HZDR~\cite{hzdrcluster}.

\appendix

\section{Geometry of epicycles}
\label{app:geometry}

Here, we follow the procedure described in~\citet{Bates13} to construct spherical epicycles with the given number of knots. The procedure can be briefly summarized as follows. For a given spherical curve $\vec{\beta} = \{h, k, l\}$ with trigonometric polynomials $h$, $k$ and $l$ satisfying $h^2+k^2+l^2 = 1$ (i.e., the curve $\vec{\beta}$ lies on a unit sphere), the epicycle~\eqref{eq:koenigs} reads
\begin{equation}\label{eq:app:int}
	\begin{split}
		x &= \frac{1}{\tau} \int_0^s \frac{l\dd k - k \dd l}{h^2 + k^2 + l^2},\\
		y &= \frac{1}{\tau} \int_0^s \frac{h\dd l - l \dd h}{h^2 + k^2 + l^2},\\
		z &= \frac{1}{\tau} \int_0^s \frac{k\dd h - h \dd k}{h^2 + k^2 + l^2}.
	\end{split}
\end{equation}
A closed curve of constant torsion can be obtained if these polynomials vanish simultaneously being integrated over $[0,2\pi]$. We can chose a spherical epicycle $\vec{\beta}$ of the symmetry $C_{|p|}$, which traces the sphere $r$ times and has $|p|$ knots characterized by two geodesic radii $\alpha_{1,2}$. The selection of the geodesic radii should be done in such a way that integrals~\eqref{eq:app:int} vanish. This determines the shape of the respective curve of the constant torsion. The parameters, used to construct epicycles in our work are shown in Table~\ref{tab:param}. Finally, we rotate the laboratory reference frame to have $\vec{\hat{z}}$ axis parallel to the epicycle's axis.

\begin{table}
	\centering
	\caption{Parameters used to build epicycles $\gamma_p^r$ from a spherical curve $\vec{\beta}$ covering the unit sphere $r$ times and having $|p|$ knots.}
	\begin{tabular}{|l|c|c|c|c|}
		\toprule
		Symmetry &  $r$ & $p$ & $\alpha_1$ & $\alpha_2$\\
		\hline
		$C_3$ &   1  &  $-3$ & $\frac{\pi}{4}$ & $\frac{1}{2}\arccos{\left(\frac{67-24\sqrt{2}}{71}\right)}$\\
		$C_4$ & 1 & $-4$ & 1 & 0.59973787219\\
		\botrule
	\end{tabular}
	\label{tab:param}
\end{table}

The shape of the curvature $\varkappa$ for the given $\sigma$ near the maximum can be analytically approximated by a Lorenzian as
\begin{equation}\label{app:lor}
	f(s) = A \; \frac{\Gamma^2}{\Gamma^2 + (s-s_0)^2} + \epsilon,
\end{equation}
with $A$ being amplitude, $\Gamma$ being half-width at half-maximum, $s_0$ being the maximum position and $\epsilon$ being offset, see Fig.~\ref{fig:first}(b). In general, $A \propto \sigma$ and $\Gamma \propto 1/|\sigma|$. The epicycle $\gamma_p^r$ can be inscribed in a circle of radius $R \propto 1/|\sigma|$.
\section{Exchange energy of a frustrated spin chain}
\label{app:frustrated}

To derive the $\sigma$-model for curvilinear spin chains with odd number of spins, we follow the procedure for AFM spin chains~\cite{Pylypovskyi21e} taking into account the periodic boundary conditions for this geometry. In this case, the exchange part of the Hamiltonian~\eqref{eq:ham} reads
\begin{equation}\label{eq:exch-ham-mn}
	\begin{aligned}
		\dfrac{\mathscr{H}_\text{x}}{JS^2} & = \dfrac{1}{2} \sum_{i=1}^{N-1} \left[ 8m_i^2 + (\Delta \vec{n}_i)^2  - (\Delta \vec{m}_i)^2  \right. \\
		& \left.+ 2(\vec{m}_i\Delta \vec{n}_{i+1} - \vec{n}_i \Delta \vec{m}_{i+1}) \right] - 2(m_N^2 + 2m_1^2) \\
		& + \vec{\mu}_0 \cdot (\vec{m}_N + \vec{m}_1) + \underbrace{\vec{\mu}_0 \cdot (\vec{n}_1 - \vec{n}_N)}_{=\delta \mathscr{H}_\text{x}/(JS^2)}.
	\end{aligned}
\end{equation}
Here, the sum over $i$ is a source of the expression~\eqref{eq:exchange} by the replacement of derivatives $\Delta f_i \to 2af'(s)$ and sums $\sum_i f_i \to 1/(2a) \int f(s)ds$ for a certain energy term $f$~\cite{Pylypovskyi21e}. The last terms outside the summation sign  appear due to symmetrization. The last two terms, which are proportional to $\vec{\mu}_0$, represent energy of the magnetic defect associated with the unpaired moment. In fact, this defect is virtual because it cannot be associated with a specific lattice site due to freedom of choice of the spin with $i = 1$ and respective shift of the origin of the reference frame. In~Eq.~\eqref{eq:exch-ham-mn}, the terms outside the sum, which are dependent on $\vec{m}_1$ and $\vec{m}_N$, are small for the case of small magnetization and under assumption that it behaves as a driving variable. 

The last term in Eq.~\eqref{eq:exch-ham-mn} is driven by the primary order parameter. Therefore, it should approach its minimal value in equilibrium. Since the transition to the continuous energy functional is possible for small gradients of $\vec{n}$ and $a \to 0$, we assume that any small amount of neighboring spins will form a checkerboard order along the certain axis and an inhomogeneity of the magnetic texture can be neglected. Then $\vec{n}_1$ and $\vec{n}_N$ should be either parallel or antiparallel. The requirement that the local spin ordering should be antiferromagnetic everywhere imposes that $\vec{\mu}_0$ is necessarily antiparallel to both, $\vec{n}_1$ and $\vec{n}_N$, thus $\vec{n}_1$ is co-aligned with $\vec{n}_N$. Therefore, the continuum N\'{e}el vector possesses the antiperiodic boundary conditions $\vec{n}(0) = - \vec{n}(L)$.

\section{Equations of state}
\label{app:eq-states}

To describe FM and AFM epicycles with the easy axis of anisotropy ($K > 0$), we use the parameterization~\eqref{eq:ea-magnetic-angles}. The polar and azimuthal angles $\Theta_\text{ea}(\xi)$ and $\Phi_\text{ea}(\xi)$, respectively, are determined by the equations 
\begin{widetext}
	\begin{equation}
		\begin{aligned}
			\Theta'' - \sin\Theta\cos\Theta \left[k_{11}(\xi) + \Phi'^2 - (1+k_{33})\cos^2\Phi + \dfrac{d_\textsc{b}(\xi)}{2}\Phi' - \dfrac{d_\textsc{b}'(\xi)}{4}\sin2\Phi \right] - k_{13}(\xi)\cos\Phi & \\
			 + \dfrac{4k_{13}(\xi)-d_\textsc{b}'(\xi)}{2}\cos^2\Theta\cos\Phi+ \dfrac{1}{2}\left[ d_\textsc{b}(\xi)\sin\Phi - 2d_\textsc{t}\cos\Phi \right] \sin^2\Theta\Phi' & = 0,\\
			 \sin^2\Theta \Phi'' + \dfrac{1}{2}\sin\Theta\cos\Theta \Bigl\lbrace[d_\textsc{b}(\xi)+4\Phi']\Theta' + [d_\textsc{b}'(\xi)-2k_{13}(\xi)]\sin\Phi\Bigr\rbrace & \\
			 - \dfrac{\sin^2\Theta}{2}\Bigl\lbrace \left[ d_\textsc{b}(\xi)\sin\Phi - 2d_\textsc{t}\cos\Phi \right]\Theta' - d_\textsc{b}'(\xi)\cos^2\Phi + (1+k_{33})\sin2\Phi \Bigr\rbrace & = 0,
		\end{aligned}
	\end{equation}
\end{widetext}
where the subscript ``ea'' is omitted for simplicity.

To describe FM and AFM epicycles with the hard axis of anisotropy ($K < 0$), we use the parameterization~\eqref{eq:ha-magnetic-angles}. The polar and azimuthal angles $\Theta_\text{ha}(\xi)$ and $\Phi_\text{ha}(\xi)$, respectively, are determined by the equations 
\begin{widetext}
	\begin{equation}
		\begin{aligned}
			\Theta'' + \sin\Theta\cos\Theta \left[ 1 -k_{33} + k_{11}(\xi)\sin^2\Phi - d_\textsc{t} \Phi' - \Phi'^2 \right] + k_{13}(\xi)\sin\Phi & \\
			+ \Bigl\lbrace d_\textsc{b}(\xi)\cos\Phi \Phi' + \left[ d_\textsc{b}'(\xi)-4k_{13}(\xi)-d_\textsc{b}\Phi' \right]\sin\Phi\Bigr\rbrace \dfrac{\sin^2\Theta}{2} & = 0,\\
			\sin^2\Theta \Phi'' + \sin\Theta\cos\Theta \left[  k_{13}(\xi)\cos\Phi + d_\textsc{t}\Theta' + 2\Theta'\Phi' - \dfrac{d_\textsc{b}'(\xi)}{2}\sin\Phi \right] & \\
			+
			 \dfrac{\sin^2\Theta}{2} \left[ d_\textsc{b}(\xi) (\sin\Phi - \cos\Phi)\Theta' + k_{11}(\xi)\sin2\Phi \right] & = 0,
		\end{aligned}
	\end{equation}
\end{widetext}
where the subscript ``ha'' is omitted for simplicity.

\section{Spin-lattice simulations}
\label{app:simulations}

To analyze magnetic spin chains numerically, we solve the Landau--Lifshitz--Gilbert equation
\begin{equation}\label{eq:llg}
	\dfrac{d\vec{\mu}_i}{dt} = \dfrac{1}{\hbar S} \vec{\mu}_i \times \dfrac{\partial \mathscr{H}}{\partial \vec{\mu}_i} + \alpha_\textsc{g} \vec{\mu}_i \times \dfrac{d\vec{\mu}_i}{dt},\quad i=\overline{1,N},
\end{equation}
using the in-house-developed spin-lattice simulation suite SLaSi~\cite{slasi} with the midpoint integration scheme. Here, $t$ is time, $\hbar$ is the reduced Planck constant and $\alpha_\textsc{g}$ is the Gilbert damping parameter. For all simulations, we chose scales between $J$, $K$ and $\mu_\textsc{b}$ in such a way to have $\ell/a = 10$. The value of magnetic length $\ell$ is calculated based on the effective anisotropy stemming from the dipolar interaction for the given sign of $J$, see Sec.~\ref{sec:sigma-model} for details. 

For simulations of magnetically soft systems, we set $\mathcal{K} = 0$. For simulations of magnetically hard systems, $\vec{H}_\text{d} \equiv 0$ is imposed. The numerical procedure to determine the ground states is as follows. By choosing one of the pre-defined initial magnetic textures, the system is relaxed by solving Eq.~\eqref{eq:llg}. The total time of relaxation for majority of simulations is $t_{\max} \approx 174 \hbar S/\mathcal{K}$ with the integration step $\Delta t \approx 0.000174 \hbar S/\mathcal{K}$. In some cases, the simulation time is extended up to $870 \hbar S/\mathcal{K}$. The sets of initial states for all studied systems include the orientation of $\vec{n}$ along $\pm \vec{\hat{x}}$, $\pm \vec{\hat{y}}$, $\pm \vec{\hat{z}}$, $\pm \vec{e}_\textsc{t}$, $\pm \vec{e}_\textsc{n}$, $\pm \vec{e}_\textsc{b}$ and several runs from different random states. To get the lowest-energy states for hard-axis AFM and frustrated epicycles, additional initial states with $\Theta_\text{ini} = \pi/2$ and piece-wise profiles of $\Phi_\text{ini}$ close to the final state are used. 

The discrete Frenet--Serret basis $\{\vec{e}_\textsc{t},\vec{e}_\textsc{n},\vec{e}_\textsc{b}\}$ is calculated based on the shape of $\vec{\gamma}$~\cite{Hu11}, which is numerically determined with a step much finer than the inter-site distance $a$. We note, that for the chosen $\ell/a$ ratio, the same procedure can be applied to the set of magnetic lattice sites while the discreteness effects are still negligibly small.

To check the critical torsion at which the ground state is changed, we performed a series of simulations of the easy-axis FM epicycles $\gamma_{-3}^1$ for different $\sigma$, where the initial states are selected to be along $\vec{e}_\textsc{t}$ and along the axis of symmetry of the epicycle. While all states remain similar visually, at $\sigma_\text{cr} = 0.377\pm 0.016$ there is a change of the spatial distribution of $\Phi_\text{ea}$. For $\sigma < \sigma_\text{cr}$, the azimuthal angle behaves qualitatively similar to the picture shown in~Fig.~\ref{fig:ea-epicycles}(b) with $\max |\Phi_\text{ea}| \sim 0.1\pi$. For $\sigma > \sigma_\text{cr}$, there is a continuous change of $\Phi$ by $6\pi$ at $0 \le \xi < X$, which is an indication of the difference in topology of magnetic states because of the helimagnetic transition due to a strong curvature-induced DMI~\cite{Sheka15a}.
 

\begin{thebibliography}{61}%
\makeatletter
\providecommand \@ifxundefined [1]{%
 \@ifx{#1\undefined}
}%
\providecommand \@ifnum [1]{%
 \ifnum #1\expandafter \@firstoftwo
 \else \expandafter \@secondoftwo
 \fi
}%
\providecommand \@ifx [1]{%
 \ifx #1\expandafter \@firstoftwo
 \else \expandafter \@secondoftwo
 \fi
}%
\providecommand \natexlab [1]{#1}%
\providecommand \enquote  [1]{``#1''}%
\providecommand \bibnamefont  [1]{#1}%
\providecommand \bibfnamefont [1]{#1}%
\providecommand \citenamefont [1]{#1}%
\providecommand \href@noop [0]{\@secondoftwo}%
\providecommand \href [0]{\begingroup \@sanitize@url \@href}%
\providecommand \@href[1]{\@@startlink{#1}\@@href}%
\providecommand \@@href[1]{\endgroup#1\@@endlink}%
\providecommand \@sanitize@url [0]{\catcode `\\12\catcode `\$12\catcode
  `\&12\catcode `\#12\catcode `\^12\catcode `\_12\catcode `\%12\relax}%
\providecommand \@@startlink[1]{}%
\providecommand \@@endlink[0]{}%
\providecommand \url  [0]{\begingroup\@sanitize@url \@url }%
\providecommand \@url [1]{\endgroup\@href {#1}{\urlprefix }}%
\providecommand \urlprefix  [0]{URL }%
\providecommand \Eprint [0]{\href }%
\providecommand \doibase [0]{https://doi.org/}%
\providecommand \selectlanguage [0]{\@gobble}%
\providecommand \bibinfo  [0]{\@secondoftwo}%
\providecommand \bibfield  [0]{\@secondoftwo}%
\providecommand \translation [1]{[#1]}%
\providecommand \BibitemOpen [0]{}%
\providecommand \bibitemStop [0]{}%
\providecommand \bibitemNoStop [0]{.\EOS\space}%
\providecommand \EOS [0]{\spacefactor3000\relax}%
\providecommand \BibitemShut  [1]{\csname bibitem#1\endcsname}%
\let\auto@bib@innerbib\@empty
\bibitem [{\citenamefont {Gentile}\ \emph {et~al.}(2022)\citenamefont
  {Gentile}, \citenamefont {Cuoco}, \citenamefont {Volkov}, \citenamefont
  {Ying}, \citenamefont {Vera-Marun}, \citenamefont {Makarov},\ and\
  \citenamefont {Ortix}}]{Gentile22}%
  \BibitemOpen
  \bibfield  {author} {\bibinfo {author} {\bibfnamefont {P.}~\bibnamefont
  {Gentile}}, \bibinfo {author} {\bibfnamefont {M.}~\bibnamefont {Cuoco}},
  \bibinfo {author} {\bibfnamefont {O.~M.}\ \bibnamefont {Volkov}}, \bibinfo
  {author} {\bibfnamefont {Z.-J.}\ \bibnamefont {Ying}}, \bibinfo {author}
  {\bibfnamefont {I.~J.}\ \bibnamefont {Vera-Marun}}, \bibinfo {author}
  {\bibfnamefont {D.}~\bibnamefont {Makarov}},\ and\ \bibinfo {author}
  {\bibfnamefont {C.}~\bibnamefont {Ortix}},\ }\bibfield  {title} {\bibinfo
  {title} {Electronic materials with nanoscale curved geometries},\ }\href
  {https://doi.org/10.1038/s41928-022-00820-z} {\bibfield  {journal} {\bibinfo
  {journal} {Nature Electronics}\ }\textbf {\bibinfo {volume} {5}},\ \bibinfo
  {pages} {551} (\bibinfo {year} {2022})}\BibitemShut {NoStop}%
\bibitem [{\citenamefont {Vozmediano}\ \emph {et~al.}(2010)\citenamefont
  {Vozmediano}, \citenamefont {Katsnelson},\ and\ \citenamefont
  {Guinea}}]{Vozmediano10}%
  \BibitemOpen
  \bibfield  {author} {\bibinfo {author} {\bibfnamefont {M.}~\bibnamefont
  {Vozmediano}}, \bibinfo {author} {\bibfnamefont {M.}~\bibnamefont
  {Katsnelson}},\ and\ \bibinfo {author} {\bibfnamefont {F.}~\bibnamefont
  {Guinea}},\ }\bibfield  {title} {\bibinfo {title} {Gauge fields in
  graphene},\ }\href
  {https://doi.org/http://dx.doi.org/10.1016/j.physrep.2010.07.003} {\bibfield
  {journal} {\bibinfo  {journal} {Physics Reports}\ }\textbf {\bibinfo {volume}
  {496}},\ \bibinfo {pages} {109} (\bibinfo {year} {2010})}\BibitemShut
  {NoStop}%
\bibitem [{\citenamefont {Grushin}\ \emph {et~al.}(2016)\citenamefont
  {Grushin}, \citenamefont {Venderbos}, \citenamefont {Vishwanath},\ and\
  \citenamefont {Ilan}}]{Grushin16}%
  \BibitemOpen
  \bibfield  {author} {\bibinfo {author} {\bibfnamefont {A.~G.}\ \bibnamefont
  {Grushin}}, \bibinfo {author} {\bibfnamefont {J.~W.~F.}\ \bibnamefont
  {Venderbos}}, \bibinfo {author} {\bibfnamefont {A.}~\bibnamefont
  {Vishwanath}},\ and\ \bibinfo {author} {\bibfnamefont {R.}~\bibnamefont
  {Ilan}},\ }\bibfield  {title} {\bibinfo {title} {Inhomogeneous weyl and dirac
  semimetals: Transport in axial magnetic fields and fermi arc surface states
  from pseudo-landau levels},\ }\href
  {https://doi.org/10.1103/physrevx.6.041046} {\bibfield  {journal} {\bibinfo
  {journal} {Physical Review X}\ }\textbf {\bibinfo {volume} {6}},\ \bibinfo
  {pages} {041046} (\bibinfo {year} {2016})}\BibitemShut {NoStop}%
\bibitem [{\citenamefont {Fomin}\ and\ \citenamefont
  {Dobrovolskiy}(2022)}]{Fomin22}%
  \BibitemOpen
  \bibfield  {author} {\bibinfo {author} {\bibfnamefont {V.~M.}\ \bibnamefont
  {Fomin}}\ and\ \bibinfo {author} {\bibfnamefont {O.~V.}\ \bibnamefont
  {Dobrovolskiy}},\ }\bibfield  {title} {\bibinfo {title} {A perspective on
  superconductivity in curved {3D} nanoarchitectures},\ }\href
  {https://doi.org/10.1063/5.0085095} {\bibfield  {journal} {\bibinfo
  {journal} {Applied Physics Letters}\ }\textbf {\bibinfo {volume} {120}},\
  \bibinfo {pages} {090501} (\bibinfo {year} {2022})}\BibitemShut {NoStop}%
\bibitem [{\citenamefont {Makarov}\ \emph {et~al.}(2022)\citenamefont
  {Makarov}, \citenamefont {Volkov}, \citenamefont {K\'{a}kay}, \citenamefont
  {Pylypovskyi}, \citenamefont {Budinsk\'{a}},\ and\ \citenamefont
  {Dobrovolskiy}}]{Makarov22}%
  \BibitemOpen
  \bibfield  {author} {\bibinfo {author} {\bibfnamefont {D.}~\bibnamefont
  {Makarov}}, \bibinfo {author} {\bibfnamefont {O.~M.}\ \bibnamefont {Volkov}},
  \bibinfo {author} {\bibfnamefont {A.}~\bibnamefont {K\'{a}kay}}, \bibinfo
  {author} {\bibfnamefont {O.~V.}\ \bibnamefont {Pylypovskyi}}, \bibinfo
  {author} {\bibfnamefont {B.}~\bibnamefont {Budinsk\'{a}}},\ and\ \bibinfo
  {author} {\bibfnamefont {O.~V.}\ \bibnamefont {Dobrovolskiy}},\ }\bibfield
  {title} {\bibinfo {title} {New dimension in magnetism and superconductivity:
  {3D} and curvilinear nanoarchitectures},\ }\href
  {https://doi.org/10.1002/adma.202101758} {\bibfield  {journal} {\bibinfo
  {journal} {Advanced Materials}\ }\textbf {\bibinfo {volume} {34}},\ \bibinfo
  {pages} {2101758} (\bibinfo {year} {2022})}\BibitemShut {NoStop}%
\bibitem [{\citenamefont {Raftrey}\ \emph {et~al.}(2022)\citenamefont
  {Raftrey}, \citenamefont {Hierro-Rodriguez}, \citenamefont
  {Fernandez-Pacheco},\ and\ \citenamefont {Fischer}}]{Raftrey22}%
  \BibitemOpen
  \bibfield  {author} {\bibinfo {author} {\bibfnamefont {D.}~\bibnamefont
  {Raftrey}}, \bibinfo {author} {\bibfnamefont {A.}~\bibnamefont
  {Hierro-Rodriguez}}, \bibinfo {author} {\bibfnamefont {A.}~\bibnamefont
  {Fernandez-Pacheco}},\ and\ \bibinfo {author} {\bibfnamefont
  {P.}~\bibnamefont {Fischer}},\ }\bibfield  {title} {\bibinfo {title} {The
  road to 3-dim nanomagnetism: {S}teep curves and architectured crosswalks},\
  }\href {https://doi.org/10.1016/j.jmmm.2022.169899} {\bibfield  {journal}
  {\bibinfo  {journal} {Journal of Magnetism and Magnetic Materials}\ ,\
  \bibinfo {pages} {169899}} (\bibinfo {year} {2022})}\BibitemShut {NoStop}%
\bibitem [{\citenamefont {Makarov}\ and\ \citenamefont
  {Sheka}(2022)}]{Makarov22a}%
  \BibitemOpen
  \bibinfo {editor} {\bibfnamefont {D.}~\bibnamefont {Makarov}}\ and\ \bibinfo
  {editor} {\bibfnamefont {D.~D.}\ \bibnamefont {Sheka}},\ eds.,\ \href
  {https://doi.org/10.1007/978-3-031-09086-8} {\emph {\bibinfo {title}
  {Curvilinear Micromagnetism}}}\ (\bibinfo  {publisher} {Springer
  International Publishing},\ \bibinfo {address} {Springer Cham},\ \bibinfo
  {year} {2022})\BibitemShut {NoStop}%
\bibitem [{\citenamefont {Sheka}\ \emph {et~al.}(2022)\citenamefont {Sheka},
  \citenamefont {Pylypovskyi}, \citenamefont {Volkov}, \citenamefont {Yershov},
  \citenamefont {Kravchuk},\ and\ \citenamefont {Makarov}}]{Sheka22}%
  \BibitemOpen
  \bibfield  {author} {\bibinfo {author} {\bibfnamefont {D.~D.}\ \bibnamefont
  {Sheka}}, \bibinfo {author} {\bibfnamefont {O.~V.}\ \bibnamefont
  {Pylypovskyi}}, \bibinfo {author} {\bibfnamefont {O.~M.}\ \bibnamefont
  {Volkov}}, \bibinfo {author} {\bibfnamefont {K.~V.}\ \bibnamefont {Yershov}},
  \bibinfo {author} {\bibfnamefont {V.~P.}\ \bibnamefont {Kravchuk}},\ and\
  \bibinfo {author} {\bibfnamefont {D.}~\bibnamefont {Makarov}},\ }\bibfield
  {title} {\bibinfo {title} {Fundamentals of curvilinear ferromagnetism:
  Statics and dynamics of geometrically curved wires and narrow ribbons},\
  }\href {https://doi.org/10.1002/smll.202105219} {\bibfield  {journal}
  {\bibinfo  {journal} {Small}\ }\textbf {\bibinfo {volume} {18}},\ \bibinfo
  {pages} {2105219} (\bibinfo {year} {2022})}\BibitemShut {NoStop}%
\bibitem [{\citenamefont {Ortix}\ and\ \citenamefont {van~den
  Brink}(2023)}]{Ortix23}%
  \BibitemOpen
  \bibfield  {author} {\bibinfo {author} {\bibfnamefont {C.}~\bibnamefont
  {Ortix}}\ and\ \bibinfo {author} {\bibfnamefont {J.}~\bibnamefont {van~den
  Brink}},\ }\bibfield  {title} {\bibinfo {title} {Magnetoelectricity induced
  by rippling of magnetic nanomembranes and wires},\ }\href
  {https://doi.org/10.1103/physrevresearch.5.l022063} {\bibfield  {journal}
  {\bibinfo  {journal} {Physical Review Research}\ }\textbf {\bibinfo {volume}
  {5}},\ \bibinfo {pages} {L022063} (\bibinfo {year} {2023})}\BibitemShut
  {NoStop}%
\bibitem [{\citenamefont {Fiebig}\ \emph {et~al.}(2016)\citenamefont {Fiebig},
  \citenamefont {Lottermoser}, \citenamefont {Meier},\ and\ \citenamefont
  {Trassin}}]{Fiebig16}%
  \BibitemOpen
  \bibfield  {author} {\bibinfo {author} {\bibfnamefont {M.}~\bibnamefont
  {Fiebig}}, \bibinfo {author} {\bibfnamefont {T.}~\bibnamefont {Lottermoser}},
  \bibinfo {author} {\bibfnamefont {D.}~\bibnamefont {Meier}},\ and\ \bibinfo
  {author} {\bibfnamefont {M.}~\bibnamefont {Trassin}},\ }\bibfield  {title}
  {\bibinfo {title} {The evolution of multiferroics},\ }\href
  {https://doi.org/10.1038/natrevmats.2016.46} {\bibfield  {journal} {\bibinfo
  {journal} {Nature Reviews Materials}\ }\textbf {\bibinfo {volume} {1}},\
  \bibinfo {pages} {16046} (\bibinfo {year} {2016})}\BibitemShut {NoStop}%
\bibitem [{\citenamefont {Phatak}\ \emph {et~al.}(2014)\citenamefont {Phatak},
  \citenamefont {Liu}, \citenamefont {Gulsoy}, \citenamefont {Schmidt},
  \citenamefont {Franke-Schubert},\ and\ \citenamefont
  {Petford-Long}}]{Phatak14}%
  \BibitemOpen
  \bibfield  {author} {\bibinfo {author} {\bibfnamefont {C.}~\bibnamefont
  {Phatak}}, \bibinfo {author} {\bibfnamefont {Y.}~\bibnamefont {Liu}},
  \bibinfo {author} {\bibfnamefont {E.~B.}\ \bibnamefont {Gulsoy}}, \bibinfo
  {author} {\bibfnamefont {D.}~\bibnamefont {Schmidt}}, \bibinfo {author}
  {\bibfnamefont {E.}~\bibnamefont {Franke-Schubert}},\ and\ \bibinfo {author}
  {\bibfnamefont {A.}~\bibnamefont {Petford-Long}},\ }\bibfield  {title}
  {\bibinfo {title} {Visualization of the magnetic structure of sculpted
  three-dimensional {C}obalt nanospirals},\ }\href
  {https://doi.org/10.1021/nl404071u} {\bibfield  {journal} {\bibinfo
  {journal} {Nano Letters}\ }\textbf {\bibinfo {volume} {14}},\ \bibinfo
  {pages} {759} (\bibinfo {year} {2014})}\BibitemShut {NoStop}%
\bibitem [{\citenamefont {Magdanz}\ \emph {et~al.}(2017)\citenamefont
  {Magdanz}, \citenamefont {Medina-S{\'{a}}nchez}, \citenamefont {Schwarz},
  \citenamefont {Xu}, \citenamefont {Elgeti},\ and\ \citenamefont
  {Schmidt}}]{Magdanz17}%
  \BibitemOpen
  \bibfield  {author} {\bibinfo {author} {\bibfnamefont {V.}~\bibnamefont
  {Magdanz}}, \bibinfo {author} {\bibfnamefont {M.}~\bibnamefont
  {Medina-S{\'{a}}nchez}}, \bibinfo {author} {\bibfnamefont {L.}~\bibnamefont
  {Schwarz}}, \bibinfo {author} {\bibfnamefont {H.}~\bibnamefont {Xu}},
  \bibinfo {author} {\bibfnamefont {J.}~\bibnamefont {Elgeti}},\ and\ \bibinfo
  {author} {\bibfnamefont {O.~G.}\ \bibnamefont {Schmidt}},\ }\bibfield
  {title} {\bibinfo {title} {Spermatozoa as functional components of robotic
  microswimmers},\ }\href {https://doi.org/10.1002/adma.201606301} {\bibfield
  {journal} {\bibinfo  {journal} {Advanced Materials}\ }\textbf {\bibinfo
  {volume} {29}},\ \bibinfo {pages} {1606301} (\bibinfo {year}
  {2017})}\BibitemShut {NoStop}%
\bibitem [{\citenamefont {Nam}\ \emph {et~al.}(2018)\citenamefont {Nam},
  \citenamefont {Samardak}, \citenamefont {Jeon}, \citenamefont {Kim},
  \citenamefont {Davydenko}, \citenamefont {Ognev}, \citenamefont {Samardak},\
  and\ \citenamefont {Kim}}]{Nam18}%
  \BibitemOpen
  \bibfield  {author} {\bibinfo {author} {\bibfnamefont {D.~Y.}\ \bibnamefont
  {Nam}}, \bibinfo {author} {\bibfnamefont {A.~Y.}\ \bibnamefont {Samardak}},
  \bibinfo {author} {\bibfnamefont {Y.~S.}\ \bibnamefont {Jeon}}, \bibinfo
  {author} {\bibfnamefont {S.~H.}\ \bibnamefont {Kim}}, \bibinfo {author}
  {\bibfnamefont {A.~V.}\ \bibnamefont {Davydenko}}, \bibinfo {author}
  {\bibfnamefont {A.~V.}\ \bibnamefont {Ognev}}, \bibinfo {author}
  {\bibfnamefont {A.~S.}\ \bibnamefont {Samardak}},\ and\ \bibinfo {author}
  {\bibfnamefont {Y.~K.}\ \bibnamefont {Kim}},\ }\bibfield  {title} {\bibinfo
  {title} {Magnetization reversal of ferromagnetic nanosprings affected by
  helical shape},\ }\href {https://doi.org/10.1039/c8nr05655b} {\bibfield
  {journal} {\bibinfo  {journal} {Nanoscale}\ }\textbf {\bibinfo {volume}
  {10}},\ \bibinfo {pages} {20405} (\bibinfo {year} {2018})}\BibitemShut
  {NoStop}%
\bibitem [{\citenamefont {Sanz-Hern{\'{a}}ndez}\ \emph
  {et~al.}(2020)\citenamefont {Sanz-Hern{\'{a}}ndez}, \citenamefont
  {Hierro-Rodriguez}, \citenamefont {Donnelly}, \citenamefont {Pablo-Navarro},
  \citenamefont {Sorrentino}, \citenamefont {Pereiro}, \citenamefont
  {Mag{\'{e}}n}, \citenamefont {McVitie}, \citenamefont {de~Teresa},
  \citenamefont {Ferrer}, \citenamefont {Fischer},\ and\ \citenamefont
  {Fern{\'{a}}ndez-Pacheco}}]{Sanz-Hernandez20}%
  \BibitemOpen
  \bibfield  {author} {\bibinfo {author} {\bibfnamefont {D.}~\bibnamefont
  {Sanz-Hern{\'{a}}ndez}}, \bibinfo {author} {\bibfnamefont {A.}~\bibnamefont
  {Hierro-Rodriguez}}, \bibinfo {author} {\bibfnamefont {C.}~\bibnamefont
  {Donnelly}}, \bibinfo {author} {\bibfnamefont {J.}~\bibnamefont
  {Pablo-Navarro}}, \bibinfo {author} {\bibfnamefont {A.}~\bibnamefont
  {Sorrentino}}, \bibinfo {author} {\bibfnamefont {E.}~\bibnamefont {Pereiro}},
  \bibinfo {author} {\bibfnamefont {C.}~\bibnamefont {Mag{\'{e}}n}}, \bibinfo
  {author} {\bibfnamefont {S.}~\bibnamefont {McVitie}}, \bibinfo {author}
  {\bibfnamefont {J.~M.}\ \bibnamefont {de~Teresa}}, \bibinfo {author}
  {\bibfnamefont {S.}~\bibnamefont {Ferrer}}, \bibinfo {author} {\bibfnamefont
  {P.}~\bibnamefont {Fischer}},\ and\ \bibinfo {author} {\bibfnamefont
  {A.}~\bibnamefont {Fern{\'{a}}ndez-Pacheco}},\ }\bibfield  {title} {\bibinfo
  {title} {Artificial double-helix for geometrical control of magnetic
  chirality},\ }\href {https://doi.org/10.1021/acsnano.0c00720} {\bibfield
  {journal} {\bibinfo  {journal} {{ACS} Nano}\ }\textbf {\bibinfo {volume}
  {14}},\ \bibinfo {pages} {8084} (\bibinfo {year} {2020})}\BibitemShut
  {NoStop}%
\bibitem [{\citenamefont {Phatak}\ \emph {et~al.}(2020)\citenamefont {Phatak},
  \citenamefont {Miller}, \citenamefont {Thompson}, \citenamefont {Gulsoy},\
  and\ \citenamefont {Petford-Long}}]{Phatak20}%
  \BibitemOpen
  \bibfield  {author} {\bibinfo {author} {\bibfnamefont {C.}~\bibnamefont
  {Phatak}}, \bibinfo {author} {\bibfnamefont {C.~S.}\ \bibnamefont {Miller}},
  \bibinfo {author} {\bibfnamefont {Z.}~\bibnamefont {Thompson}}, \bibinfo
  {author} {\bibfnamefont {E.~B.}\ \bibnamefont {Gulsoy}},\ and\ \bibinfo
  {author} {\bibfnamefont {A.~K.}\ \bibnamefont {Petford-Long}},\ }\bibfield
  {title} {\bibinfo {title} {Curved three-dimensional cobalt nanohelices for
  use in domain wall device applications},\ }\href
  {https://doi.org/10.1021/acsanm.0c01228} {\bibfield  {journal} {\bibinfo
  {journal} {{ACS} Applied Nano Materials}\ }\textbf {\bibinfo {volume} {3}},\
  \bibinfo {pages} {6009} (\bibinfo {year} {2020})}\BibitemShut {NoStop}%
\bibitem [{\citenamefont {Fullerton}\ \emph {et~al.}(2024)\citenamefont
  {Fullerton}, \citenamefont {McCray}, \citenamefont {Petford-Long},\ and\
  \citenamefont {Phatak}}]{Fullerton24}%
  \BibitemOpen
  \bibfield  {author} {\bibinfo {author} {\bibfnamefont {J.}~\bibnamefont
  {Fullerton}}, \bibinfo {author} {\bibfnamefont {A.~R.~C.}\ \bibnamefont
  {McCray}}, \bibinfo {author} {\bibfnamefont {A.~K.}\ \bibnamefont
  {Petford-Long}},\ and\ \bibinfo {author} {\bibfnamefont {C.}~\bibnamefont
  {Phatak}},\ }\bibfield  {title} {\bibinfo {title} {Understanding the effect
  of curvature on the magnetization reversal of three-dimensional
  nanohelices},\ }\bibfield  {journal} {\bibinfo  {journal} {Nano Letters}\
  }\href {https://doi.org/10.1021/acs.nanolett.3c04172}
  {10.1021/acs.nanolett.3c04172} (\bibinfo {year} {2024})\BibitemShut {NoStop}%
\bibitem [{\citenamefont {Sheka}\ \emph
  {et~al.}(2015{\natexlab{a}})\citenamefont {Sheka}, \citenamefont {Kravchuk},
  \citenamefont {Yershov},\ and\ \citenamefont {Gaididei}}]{Sheka15a}%
  \BibitemOpen
  \bibfield  {author} {\bibinfo {author} {\bibfnamefont {D.~D.}\ \bibnamefont
  {Sheka}}, \bibinfo {author} {\bibfnamefont {V.~P.}\ \bibnamefont {Kravchuk}},
  \bibinfo {author} {\bibfnamefont {K.~V.}\ \bibnamefont {Yershov}},\ and\
  \bibinfo {author} {\bibfnamefont {Y.}~\bibnamefont {Gaididei}},\ }\bibfield
  {title} {\bibinfo {title} {Torsion-induced effects in magnetic nanowires},\
  }\href {https://doi.org/10.1103/PhysRevB.92.054417} {\bibfield  {journal}
  {\bibinfo  {journal} {Physical Review B}\ }\textbf {\bibinfo {volume} {92}},\
  \bibinfo {pages} {054417} (\bibinfo {year} {2015}{\natexlab{a}})}\BibitemShut
  {NoStop}%
\bibitem [{\citenamefont {Pylypovskyi}\ \emph {et~al.}(2016)\citenamefont
  {Pylypovskyi}, \citenamefont {Sheka}, \citenamefont {Kravchuk}, \citenamefont
  {Yershov}, \citenamefont {Makarov},\ and\ \citenamefont
  {Gaididei}}]{Pylypovskyi16}%
  \BibitemOpen
  \bibfield  {author} {\bibinfo {author} {\bibfnamefont {O.~V.}\ \bibnamefont
  {Pylypovskyi}}, \bibinfo {author} {\bibfnamefont {D.~D.}\ \bibnamefont
  {Sheka}}, \bibinfo {author} {\bibfnamefont {V.~P.}\ \bibnamefont {Kravchuk}},
  \bibinfo {author} {\bibfnamefont {K.~V.}\ \bibnamefont {Yershov}}, \bibinfo
  {author} {\bibfnamefont {D.}~\bibnamefont {Makarov}},\ and\ \bibinfo {author}
  {\bibfnamefont {Y.}~\bibnamefont {Gaididei}},\ }\bibfield  {title} {\bibinfo
  {title} {Rashba torque driven domain wall motion in magnetic helices},\
  }\href {https://doi.org/10.1038/srep23316} {\bibfield  {journal} {\bibinfo
  {journal} {Scientific Reports}\ }\textbf {\bibinfo {volume} {6}},\ \bibinfo
  {pages} {23316} (\bibinfo {year} {2016})}\BibitemShut {NoStop}%
\bibitem [{\citenamefont {Volkov}\ \emph {et~al.}(2018)\citenamefont {Volkov},
  \citenamefont {Sheka}, \citenamefont {Gaididei}, \citenamefont {Kravchuk},
  \citenamefont {R\"o{\ss}ler}, \citenamefont {Fassbender},\ and\ \citenamefont
  {Makarov}}]{Volkov18}%
  \BibitemOpen
  \bibfield  {author} {\bibinfo {author} {\bibfnamefont {O.~M.}\ \bibnamefont
  {Volkov}}, \bibinfo {author} {\bibfnamefont {D.~D.}\ \bibnamefont {Sheka}},
  \bibinfo {author} {\bibfnamefont {Y.}~\bibnamefont {Gaididei}}, \bibinfo
  {author} {\bibfnamefont {V.~P.}\ \bibnamefont {Kravchuk}}, \bibinfo {author}
  {\bibfnamefont {U.~K.}\ \bibnamefont {R\"o{\ss}ler}}, \bibinfo {author}
  {\bibfnamefont {J.}~\bibnamefont {Fassbender}},\ and\ \bibinfo {author}
  {\bibfnamefont {D.}~\bibnamefont {Makarov}},\ }\bibfield  {title} {\bibinfo
  {title} {{Mesoscale Dzyaloshinskii-Moriya interaction: geometrical tailoring
  of the magnetochirality}},\ }\href
  {https://doi.org/10.1038/s41598-017-18835-4} {\bibfield  {journal} {\bibinfo
  {journal} {Scientific Reports}\ }\textbf {\bibinfo {volume} {8}},\ \bibinfo
  {pages} {866} (\bibinfo {year} {2018})}\BibitemShut {NoStop}%
\bibitem [{\citenamefont {Volkov}\ \emph
  {et~al.}(2019{\natexlab{a}})\citenamefont {Volkov}, \citenamefont {Rossler},
  \citenamefont {Fassbender},\ and\ \citenamefont {Makarov}}]{Volkov19b}%
  \BibitemOpen
  \bibfield  {author} {\bibinfo {author} {\bibfnamefont {O.}~\bibnamefont
  {Volkov}}, \bibinfo {author} {\bibfnamefont {U.~K.}\ \bibnamefont {Rossler}},
  \bibinfo {author} {\bibfnamefont {J.}~\bibnamefont {Fassbender}},\ and\
  \bibinfo {author} {\bibfnamefont {D.}~\bibnamefont {Makarov}},\ }\bibfield
  {title} {\bibinfo {title} {Concept of artificial magnetoelectric materials
  via geometrically controlling curvilinear helimagnets},\ }\href
  {https://doi.org/10.1088/1361-6463/ab2368} {\bibfield  {journal} {\bibinfo
  {journal} {Journal of Physics D: Applied Physics}\ }\textbf {\bibinfo
  {volume} {52}},\ \bibinfo {pages} {345001} (\bibinfo {year}
  {2019}{\natexlab{a}})}\BibitemShut {NoStop}%
\bibitem [{\citenamefont {Salamone}\ \emph {et~al.}(2024)\citenamefont
  {Salamone}, \citenamefont {Hugdal}, \citenamefont {Amundsen},\ and\
  \citenamefont {Jacobsen}}]{Salamone24a}%
  \BibitemOpen
  \bibfield  {author} {\bibinfo {author} {\bibfnamefont {T.}~\bibnamefont
  {Salamone}}, \bibinfo {author} {\bibfnamefont {H.~G.}\ \bibnamefont
  {Hugdal}}, \bibinfo {author} {\bibfnamefont {M.}~\bibnamefont {Amundsen}},\
  and\ \bibinfo {author} {\bibfnamefont {S.~H.}\ \bibnamefont {Jacobsen}},\
  }\bibfield  {title} {\bibinfo {title} {Electrical control of superconducting
  spin valves using ferromagnetic helices},\ }\bibfield  {journal} {\bibinfo
  {journal} {Applied Physics Letters}\ }\textbf {\bibinfo {volume} {125}},\
  \href {https://doi.org/10.1063/5.0210691} {10.1063/5.0210691} (\bibinfo
  {year} {2024}),\ \bibinfo {note} {the geometrical properties of a helical
  ferromagnet are shown theoretically to control the critical temperature of a
  proximity-coupled superconductor. Using the Usadel equation for diffusive
  spin transport, we provide self-consistent analysis of how curvature and
  torsion modulate the proximity effect. When the helix is attached to a
  piezoelectric actuator, the pitch of the helix—and hence the
  superconducting transition—can be controlled electrically.}\BibitemShut
  {Stop}%
\bibitem [{\citenamefont {Das~Gupta}\ and\ \citenamefont
  {Maiti}(2022)}]{DasGupta22}%
  \BibitemOpen
  \bibfield  {author} {\bibinfo {author} {\bibfnamefont {D.}~\bibnamefont
  {Das~Gupta}}\ and\ \bibinfo {author} {\bibfnamefont {S.~K.}\ \bibnamefont
  {Maiti}},\ }\bibfield  {title} {\bibinfo {title} {Antiferromagnetic helix as
  an efficient spin polarizer: Interplay between electric field and
  higher-order hopping},\ }\href {https://doi.org/10.1103/physrevb.106.125420}
  {\bibfield  {journal} {\bibinfo  {journal} {Physical Review B}\ }\textbf
  {\bibinfo {volume} {106}},\ \bibinfo {pages} {125420} (\bibinfo {year}
  {2022})}\BibitemShut {NoStop}%
\bibitem [{\citenamefont {Mondal}\ \emph {et~al.}(2023)\citenamefont {Mondal},
  \citenamefont {Ganguly},\ and\ \citenamefont {Maiti}}]{Mondal23}%
  \BibitemOpen
  \bibfield  {author} {\bibinfo {author} {\bibfnamefont {K.}~\bibnamefont
  {Mondal}}, \bibinfo {author} {\bibfnamefont {S.}~\bibnamefont {Ganguly}},\
  and\ \bibinfo {author} {\bibfnamefont {S.~K.}\ \bibnamefont {Maiti}},\
  }\bibfield  {title} {\bibinfo {title} {Thermoelectric phenomena in an
  antiferromagnetic helix: Role of electric field},\ }\href
  {https://doi.org/10.1103/physrevb.108.195401} {\bibfield  {journal} {\bibinfo
   {journal} {Physical Review B}\ }\textbf {\bibinfo {volume} {108}},\ \bibinfo
  {pages} {195401} (\bibinfo {year} {2023})}\BibitemShut {NoStop}%
\bibitem [{\citenamefont {Das~Gupta}\ and\ \citenamefont
  {Maiti}(2023)}]{DasGupta23}%
  \BibitemOpen
  \bibfield  {author} {\bibinfo {author} {\bibfnamefont {D.}~\bibnamefont
  {Das~Gupta}}\ and\ \bibinfo {author} {\bibfnamefont {S.~K.}\ \bibnamefont
  {Maiti}},\ }\bibfield  {title} {\bibinfo {title} {Spin current rectification
  in a helical magnetic system with vanishing net magnetization},\ }\href
  {https://doi.org/10.1016/j.aop.2023.169343} {\bibfield  {journal} {\bibinfo
  {journal} {Annals of Physics}\ ,\ \bibinfo {pages} {169343}} (\bibinfo {year}
  {2023})}\BibitemShut {NoStop}%
\bibitem [{\citenamefont {Yershov}\ \emph {et~al.}(2016)\citenamefont
  {Yershov}, \citenamefont {Kravchuk}, \citenamefont {Sheka},\ and\
  \citenamefont {Gaididei}}]{Yershov16}%
  \BibitemOpen
  \bibfield  {author} {\bibinfo {author} {\bibfnamefont {K.~V.}\ \bibnamefont
  {Yershov}}, \bibinfo {author} {\bibfnamefont {V.~P.}\ \bibnamefont
  {Kravchuk}}, \bibinfo {author} {\bibfnamefont {D.~D.}\ \bibnamefont
  {Sheka}},\ and\ \bibinfo {author} {\bibfnamefont {Y.}~\bibnamefont
  {Gaididei}},\ }\bibfield  {title} {\bibinfo {title} {Curvature and torsion
  effects in spin-current driven domain wall motion},\ }\href
  {https://doi.org/10.1103/PhysRevB.93.094418} {\bibfield  {journal} {\bibinfo
  {journal} {Physical Review B}\ }\textbf {\bibinfo {volume} {93}},\ \bibinfo
  {pages} {094418} (\bibinfo {year} {2016})}\BibitemShut {NoStop}%
\bibitem [{\citenamefont {Moreno}\ \emph {et~al.}(2017)\citenamefont {Moreno},
  \citenamefont {Carvalho-Santos}, \citenamefont {Espejo}, \citenamefont
  {Laroze}, \citenamefont {Chubykalo-Fesenko},\ and\ \citenamefont
  {Altbir}}]{Moreno17}%
  \BibitemOpen
  \bibfield  {author} {\bibinfo {author} {\bibfnamefont {R.}~\bibnamefont
  {Moreno}}, \bibinfo {author} {\bibfnamefont {V.~L.}\ \bibnamefont
  {Carvalho-Santos}}, \bibinfo {author} {\bibfnamefont {A.~P.}\ \bibnamefont
  {Espejo}}, \bibinfo {author} {\bibfnamefont {D.}~\bibnamefont {Laroze}},
  \bibinfo {author} {\bibfnamefont {O.}~\bibnamefont {Chubykalo-Fesenko}},\
  and\ \bibinfo {author} {\bibfnamefont {D.}~\bibnamefont {Altbir}},\
  }\bibfield  {title} {\bibinfo {title} {Oscillatory behavior of the domain
  wall dynamics in a curved cylindrical magnetic nanowire},\ }\href
  {https://doi.org/10.1103/physrevb.96.184401} {\bibfield  {journal} {\bibinfo
  {journal} {Physical Review B}\ }\textbf {\bibinfo {volume} {96}},\ \bibinfo
  {pages} {184401} (\bibinfo {year} {2017})}\BibitemShut {NoStop}%
\bibitem [{\citenamefont {Volkov}\ \emph
  {et~al.}(2019{\natexlab{b}})\citenamefont {Volkov}, \citenamefont {K\'akay},
  \citenamefont {Kronast}, \citenamefont {M\"onch}, \citenamefont {Mawass},
  \citenamefont {Fassbender},\ and\ \citenamefont {Makarov}}]{Volkov19c}%
  \BibitemOpen
  \bibfield  {author} {\bibinfo {author} {\bibfnamefont {O.~M.}\ \bibnamefont
  {Volkov}}, \bibinfo {author} {\bibfnamefont {A.}~\bibnamefont {K\'akay}},
  \bibinfo {author} {\bibfnamefont {F.}~\bibnamefont {Kronast}}, \bibinfo
  {author} {\bibfnamefont {I.}~\bibnamefont {M\"onch}}, \bibinfo {author}
  {\bibfnamefont {M.-A.}\ \bibnamefont {Mawass}}, \bibinfo {author}
  {\bibfnamefont {J.}~\bibnamefont {Fassbender}},\ and\ \bibinfo {author}
  {\bibfnamefont {D.}~\bibnamefont {Makarov}},\ }\bibfield  {title} {\bibinfo
  {title} {Experimental observation of exchange-driven chiral effects in
  curvilinear magnetism},\ }\href
  {https://doi.org/10.1103/PhysRevLett.123.077201} {\bibfield  {journal}
  {\bibinfo  {journal} {Physical Review Letters}\ }\textbf {\bibinfo {volume}
  {123}},\ \bibinfo {pages} {077201} (\bibinfo {year}
  {2019}{\natexlab{b}})}\BibitemShut {NoStop}%
\bibitem [{\citenamefont {Yershov}(2022)}]{Yershov22a}%
  \BibitemOpen
  \bibfield  {author} {\bibinfo {author} {\bibfnamefont {K.~V.}\ \bibnamefont
  {Yershov}},\ }\bibfield  {title} {\bibinfo {title} {Dynamics of domain walls
  in curved antiferromagnetic wires},\ }\href
  {https://doi.org/10.1103/physrevb.105.064407} {\bibfield  {journal} {\bibinfo
   {journal} {Physical Review B}\ }\textbf {\bibinfo {volume} {105}},\ \bibinfo
  {pages} {064407} (\bibinfo {year} {2022})}\BibitemShut {NoStop}%
\bibitem [{\citenamefont {Bittencourt}\ \emph {et~al.}(2024)\citenamefont
  {Bittencourt}, \citenamefont {Carvalho-Santos}, \citenamefont
  {Chubykalo-Fesenko}, \citenamefont {Altbir},\ and\ \citenamefont
  {Moreno}}]{Bittencourt24a}%
  \BibitemOpen
  \bibfield  {author} {\bibinfo {author} {\bibfnamefont {G.~H.~R.}\
  \bibnamefont {Bittencourt}}, \bibinfo {author} {\bibfnamefont {V.~L.}\
  \bibnamefont {Carvalho-Santos}}, \bibinfo {author} {\bibfnamefont
  {O.}~\bibnamefont {Chubykalo-Fesenko}}, \bibinfo {author} {\bibfnamefont
  {D.}~\bibnamefont {Altbir}},\ and\ \bibinfo {author} {\bibfnamefont
  {R.}~\bibnamefont {Moreno}},\ }\bibfield  {title} {\bibinfo {title} {Dynamics
  of chiral domain walls in bent cylindrical magnetic nanowires},\ }\bibfield
  {journal} {\bibinfo  {journal} {Journal of Applied Physics}\ }\textbf
  {\bibinfo {volume} {135}},\ \href {https://doi.org/10.1063/5.0188985}
  {10.1063/5.0188985} (\bibinfo {year} {2024})\BibitemShut {NoStop}%
\bibitem [{\citenamefont {Yershov}\ \emph {et~al.}(2018)\citenamefont
  {Yershov}, \citenamefont {Kravchuk}, \citenamefont {Sheka}, \citenamefont
  {Pylypovskyi}, \citenamefont {Makarov},\ and\ \citenamefont
  {Gaididei}}]{Yershov18}%
  \BibitemOpen
  \bibfield  {author} {\bibinfo {author} {\bibfnamefont {K.~V.}\ \bibnamefont
  {Yershov}}, \bibinfo {author} {\bibfnamefont {V.~P.}\ \bibnamefont
  {Kravchuk}}, \bibinfo {author} {\bibfnamefont {D.~D.}\ \bibnamefont {Sheka}},
  \bibinfo {author} {\bibfnamefont {O.~V.}\ \bibnamefont {Pylypovskyi}},
  \bibinfo {author} {\bibfnamefont {D.}~\bibnamefont {Makarov}},\ and\ \bibinfo
  {author} {\bibfnamefont {Y.}~\bibnamefont {Gaididei}},\ }\bibfield  {title}
  {\bibinfo {title} {Geometry-induced motion of magnetic domain walls in curved
  nanostripes},\ }\href {https://doi.org/10.1103/physrevb.98.060409} {\bibfield
   {journal} {\bibinfo  {journal} {Physical Review B}\ }\textbf {\bibinfo
  {volume} {98}},\ \bibinfo {pages} {060409(R)} (\bibinfo {year}
  {2018})}\BibitemShut {NoStop}%
\bibitem [{\citenamefont {Bittencourt}\ \emph {et~al.}(2022)\citenamefont
  {Bittencourt}, \citenamefont {Chubykalo-Fesenko}, \citenamefont {Altbir},
  \citenamefont {Carvalho-Santos},\ and\ \citenamefont
  {Moreno}}]{Bittencourt22}%
  \BibitemOpen
  \bibfield  {author} {\bibinfo {author} {\bibfnamefont {G.~H.~R.}\
  \bibnamefont {Bittencourt}}, \bibinfo {author} {\bibfnamefont
  {O.}~\bibnamefont {Chubykalo-Fesenko}}, \bibinfo {author} {\bibfnamefont
  {D.}~\bibnamefont {Altbir}}, \bibinfo {author} {\bibfnamefont {V.~L.}\
  \bibnamefont {Carvalho-Santos}},\ and\ \bibinfo {author} {\bibfnamefont
  {R.}~\bibnamefont {Moreno}},\ }\bibfield  {title} {\bibinfo {title} {Area law
  for magnetic domain walls in bent cylindrical nanowires},\ }\href
  {https://doi.org/10.1103/physrevb.106.094410} {\bibfield  {journal} {\bibinfo
   {journal} {Physical Review B}\ }\textbf {\bibinfo {volume} {106}},\ \bibinfo
  {pages} {094410} (\bibinfo {year} {2022})}\BibitemShut {NoStop}%
\bibitem [{\citenamefont {Zhao}\ \emph {et~al.}(2023)\citenamefont {Zhao},
  \citenamefont {Cheng},\ and\ \citenamefont {Liu}}]{Zhao23}%
  \BibitemOpen
  \bibfield  {author} {\bibinfo {author} {\bibfnamefont {H.}~\bibnamefont
  {Zhao}}, \bibinfo {author} {\bibfnamefont {R.}~\bibnamefont {Cheng}},\ and\
  \bibinfo {author} {\bibfnamefont {Q.-H.}\ \bibnamefont {Liu}},\ }\bibfield
  {title} {\bibinfo {title} {Long range motion of domain wall in
  antiferromagnetic {3D} curved nanowire},\ }\href
  {https://doi.org/10.1016/j.rinp.2023.106848} {\bibfield  {journal} {\bibinfo
  {journal} {Results in Physics}\ }\textbf {\bibinfo {volume} {52}},\ \bibinfo
  {pages} {106848} (\bibinfo {year} {2023})}\BibitemShut {NoStop}%
\bibitem [{\citenamefont {Askey}\ \emph {et~al.}(2024)\citenamefont {Askey},
  \citenamefont {Hunt}, \citenamefont {Payne}, \citenamefont {van~den Berg},
  \citenamefont {Pitsios}, \citenamefont {Hejazi}, \citenamefont {Langbein},\
  and\ \citenamefont {Ladak}}]{Askey24}%
  \BibitemOpen
  \bibfield  {author} {\bibinfo {author} {\bibfnamefont {J.}~\bibnamefont
  {Askey}}, \bibinfo {author} {\bibfnamefont {M.~O.}\ \bibnamefont {Hunt}},
  \bibinfo {author} {\bibfnamefont {L.}~\bibnamefont {Payne}}, \bibinfo
  {author} {\bibfnamefont {A.}~\bibnamefont {van~den Berg}}, \bibinfo {author}
  {\bibfnamefont {I.}~\bibnamefont {Pitsios}}, \bibinfo {author} {\bibfnamefont
  {A.}~\bibnamefont {Hejazi}}, \bibinfo {author} {\bibfnamefont
  {W.}~\bibnamefont {Langbein}},\ and\ \bibinfo {author} {\bibfnamefont
  {S.}~\bibnamefont {Ladak}},\ }\bibfield  {title} {\bibinfo {title} {Direct
  visualization of domain wall pinning in sub-100nm 3d magnetic nanowires with
  cross-sectional curvature},\ }\href {https://arxiv.org/abs/2403.04411}
  {\bibfield  {journal} {\bibinfo  {journal} {ArXiv e-prints}\ } (\bibinfo
  {year} {2024})}\BibitemShut {NoStop}%
\bibitem [{\citenamefont {Bogani}\ and\ \citenamefont
  {Wernsdorfer}(2008)}]{Bogani08}%
  \BibitemOpen
  \bibfield  {author} {\bibinfo {author} {\bibfnamefont {L.}~\bibnamefont
  {Bogani}}\ and\ \bibinfo {author} {\bibfnamefont {W.}~\bibnamefont
  {Wernsdorfer}},\ }\bibfield  {title} {\bibinfo {title} {Molecular spintronics
  using single-molecule magnets},\ }\href {https://doi.org/10.1038/nmat2133}
  {\bibfield  {journal} {\bibinfo  {journal} {Nature Materials}\ }\textbf
  {\bibinfo {volume} {7}},\ \bibinfo {pages} {179} (\bibinfo {year}
  {2008})}\BibitemShut {NoStop}%
\bibitem [{\citenamefont {Ungur}\ \emph {et~al.}(2014)\citenamefont {Ungur},
  \citenamefont {Lin}, \citenamefont {Tang},\ and\ \citenamefont
  {Chibotaru}}]{Ungur14}%
  \BibitemOpen
  \bibfield  {author} {\bibinfo {author} {\bibfnamefont {L.}~\bibnamefont
  {Ungur}}, \bibinfo {author} {\bibfnamefont {S.-Y.}\ \bibnamefont {Lin}},
  \bibinfo {author} {\bibfnamefont {J.}~\bibnamefont {Tang}},\ and\ \bibinfo
  {author} {\bibfnamefont {L.~F.}\ \bibnamefont {Chibotaru}},\ }\bibfield
  {title} {\bibinfo {title} {Single-molecule toroics in ising-type lanthanide
  molecular clusters},\ }\href {https://doi.org/10.1039/c4cs00095a} {\bibfield
  {journal} {\bibinfo  {journal} {Chem. Soc. Rev.}\ }\textbf {\bibinfo {volume}
  {43}},\ \bibinfo {pages} {6894} (\bibinfo {year} {2014})}\BibitemShut
  {NoStop}%
\bibitem [{\citenamefont {Murray}(2022)}]{Murray22}%
  \BibitemOpen
  \bibinfo {editor} {\bibfnamefont {K.}~\bibnamefont {Murray}},\ ed.,\ \href
  {https://doi.org/10.1007/978-3-031-11709-1} {\emph {\bibinfo {title} {Single
  Molecule Toroics}}}\ (\bibinfo  {publisher} {Springer International
  Publishing},\ \bibinfo {address} {Cham},\ \bibinfo {year} {2022})\BibitemShut
  {NoStop}%
\bibitem [{\citenamefont {Pylypovskyi}\ \emph {et~al.}(2015)\citenamefont
  {Pylypovskyi}, \citenamefont {Kravchuk}, \citenamefont {Sheka}, \citenamefont
  {Makarov}, \citenamefont {Schmidt},\ and\ \citenamefont
  {Gaididei}}]{Pylypovskyi15b}%
  \BibitemOpen
  \bibfield  {author} {\bibinfo {author} {\bibfnamefont {O.~V.}\ \bibnamefont
  {Pylypovskyi}}, \bibinfo {author} {\bibfnamefont {V.~P.}\ \bibnamefont
  {Kravchuk}}, \bibinfo {author} {\bibfnamefont {D.~D.}\ \bibnamefont {Sheka}},
  \bibinfo {author} {\bibfnamefont {D.}~\bibnamefont {Makarov}}, \bibinfo
  {author} {\bibfnamefont {O.~G.}\ \bibnamefont {Schmidt}},\ and\ \bibinfo
  {author} {\bibfnamefont {Y.}~\bibnamefont {Gaididei}},\ }\bibfield  {title}
  {\bibinfo {title} {Coupling of chiralities in spin and physical spaces: {T}he
  {M}\"obius ring as a case study},\ }\href
  {https://doi.org/10.1103/PhysRevLett.114.197204} {\bibfield  {journal}
  {\bibinfo  {journal} {Physical Review Letters}\ }\textbf {\bibinfo {volume}
  {114}},\ \bibinfo {pages} {197204} (\bibinfo {year} {2015})}\BibitemShut
  {NoStop}%
\bibitem [{\citenamefont {Gaididei}\ \emph {et~al.}(2017)\citenamefont
  {Gaididei}, \citenamefont {Goussev}, \citenamefont {Kravchuk}, \citenamefont
  {Pylypovskyi}, \citenamefont {Robbins}, \citenamefont {Sheka}, \citenamefont
  {Slastikov},\ and\ \citenamefont {Vasylkevych}}]{Gaididei17}%
  \BibitemOpen
  \bibfield  {author} {\bibinfo {author} {\bibfnamefont {Y.}~\bibnamefont
  {Gaididei}}, \bibinfo {author} {\bibfnamefont {A.}~\bibnamefont {Goussev}},
  \bibinfo {author} {\bibfnamefont {V.~P.}\ \bibnamefont {Kravchuk}}, \bibinfo
  {author} {\bibfnamefont {O.~V.}\ \bibnamefont {Pylypovskyi}}, \bibinfo
  {author} {\bibfnamefont {J.~M.}\ \bibnamefont {Robbins}}, \bibinfo {author}
  {\bibfnamefont {D.}~\bibnamefont {Sheka}}, \bibinfo {author} {\bibfnamefont
  {V.}~\bibnamefont {Slastikov}},\ and\ \bibinfo {author} {\bibfnamefont
  {S.}~\bibnamefont {Vasylkevych}},\ }\bibfield  {title} {\bibinfo {title}
  {Magnetization in narrow ribbons: curvature effects},\ }\href
  {https://doi.org/10.1088/1751-8121/aa8179} {\bibfield  {journal} {\bibinfo
  {journal} {Journal of Physics A: Mathematical and Theoretical}\ }\textbf
  {\bibinfo {volume} {50}},\ \bibinfo {pages} {385401} (\bibinfo {year}
  {2017})}\BibitemShut {NoStop}%
\bibitem [{\citenamefont {Calini}\ and\ \citenamefont {Ivey}(1998)}]{Calini98}%
  \BibitemOpen
  \bibfield  {author} {\bibinfo {author} {\bibfnamefont {A.}~\bibnamefont
  {Calini}}\ and\ \bibinfo {author} {\bibfnamefont {T.}~\bibnamefont {Ivey}},\
  }\bibfield  {title} {\bibinfo {title} {B\"{a}cklund transformations and knots
  of constant torsion},\ }\href {https://doi.org/10.1142/s0218216598000383}
  {\bibfield  {journal} {\bibinfo  {journal} {Journal of Knot Theory and Its
  Ramifications}\ }\textbf {\bibinfo {volume} {07}},\ \bibinfo {pages} {719}
  (\bibinfo {year} {1998})}\BibitemShut {NoStop}%
\bibitem [{\citenamefont {Monterde}(2009)}]{Monterde09}%
  \BibitemOpen
  \bibfield  {author} {\bibinfo {author} {\bibfnamefont {J.}~\bibnamefont
  {Monterde}},\ }\bibfield  {title} {\bibinfo {title} {Salkowski curves
  revisited: {A} family of curves with constant curvature and non-constant
  torsion},\ }\href {https://doi.org/10.1016/j.cagd.2008.10.002} {\bibfield
  {journal} {\bibinfo  {journal} {Computer Aided Geometric Design}\ }\textbf
  {\bibinfo {volume} {26}},\ \bibinfo {pages} {271} (\bibinfo {year}
  {2009})}\BibitemShut {NoStop}%
\bibitem [{\citenamefont {Koenigs}(1887)}]{Koenigs1887}%
  \BibitemOpen
  \bibfield  {author} {\bibinfo {author} {\bibfnamefont {G.}~\bibnamefont
  {Koenigs}},\ }\bibfield  {title} {\bibinfo {title} {Sur la forme des courbes
  \`{a} torsion constante},\ }\href {https://doi.org/10.5802/afst.7} {\bibfield
   {journal} {\bibinfo  {journal} {Annales de la facult\'{e} des sciences de
  Toulouse Math\'{e}matiques}\ }\textbf {\bibinfo {volume} {1}},\ \bibinfo
  {pages} {1} (\bibinfo {year} {1887})}\BibitemShut {NoStop}%
\bibitem [{\citenamefont {Weiner}(1977)}]{Weiner77}%
  \BibitemOpen
  \bibfield  {author} {\bibinfo {author} {\bibfnamefont {J.~L.}\ \bibnamefont
  {Weiner}},\ }\bibfield  {title} {\bibinfo {title} {Closed curves of constant
  torsion. {II}},\ }\href {https://doi.org/10.1090/s0002-9939-1977-0461385-0}
  {\bibfield  {journal} {\bibinfo  {journal} {Proceedings of the American
  Mathematical Society}\ }\textbf {\bibinfo {volume} {67}},\ \bibinfo {pages}
  {306} (\bibinfo {year} {1977})}\BibitemShut {NoStop}%
\bibitem [{\citenamefont {Bates}\ and\ \citenamefont {Melko}(2013)}]{Bates13}%
  \BibitemOpen
  \bibfield  {author} {\bibinfo {author} {\bibfnamefont {L.~M.}\ \bibnamefont
  {Bates}}\ and\ \bibinfo {author} {\bibfnamefont {O.~M.}\ \bibnamefont
  {Melko}},\ }\bibfield  {title} {\bibinfo {title} {On curves of constant
  torsion {I}},\ }\href {https://doi.org/10.1007/s00022-013-0166-2} {\bibfield
  {journal} {\bibinfo  {journal} {Journal of Geometry}\ }\textbf {\bibinfo
  {volume} {104}},\ \bibinfo {pages} {213} (\bibinfo {year}
  {2013})}\BibitemShut {NoStop}%
\bibitem [{\citenamefont {Pylypovskyi}\ \emph {et~al.}(2021)\citenamefont
  {Pylypovskyi}, \citenamefont {Borysenko}, \citenamefont {Fassbender},
  \citenamefont {Sheka},\ and\ \citenamefont {Makarov}}]{Pylypovskyi21e}%
  \BibitemOpen
  \bibfield  {author} {\bibinfo {author} {\bibfnamefont {O.~V.}\ \bibnamefont
  {Pylypovskyi}}, \bibinfo {author} {\bibfnamefont {Y.~A.}\ \bibnamefont
  {Borysenko}}, \bibinfo {author} {\bibfnamefont {J.}~\bibnamefont
  {Fassbender}}, \bibinfo {author} {\bibfnamefont {D.~D.}\ \bibnamefont
  {Sheka}},\ and\ \bibinfo {author} {\bibfnamefont {D.}~\bibnamefont
  {Makarov}},\ }\bibfield  {title} {\bibinfo {title} {Curvature-driven
  homogeneous {D}zyaloshinskii--{M}oriya interaction and emergent weak
  ferromagnetism in anisotropic antiferromagnetic spin chains},\ }\href
  {https://doi.org/10.1063/5.0048823} {\bibfield  {journal} {\bibinfo
  {journal} {Applied Physics Letters}\ }\textbf {\bibinfo {volume} {118}},\
  \bibinfo {pages} {182405} (\bibinfo {year} {2021})}\BibitemShut {NoStop}%
\bibitem [{\citenamefont {Pylypovskyi}\ \emph {et~al.}(2020)\citenamefont
  {Pylypovskyi}, \citenamefont {Kononenko}, \citenamefont {Yershov},
  \citenamefont {R{\"{o}}{\ss}ler}, \citenamefont {Tomilo}, \citenamefont
  {Fassbender}, \citenamefont {van~den Brink}, \citenamefont {Makarov},\ and\
  \citenamefont {Sheka}}]{Pylypovskyi20}%
  \BibitemOpen
  \bibfield  {author} {\bibinfo {author} {\bibfnamefont {O.~V.}\ \bibnamefont
  {Pylypovskyi}}, \bibinfo {author} {\bibfnamefont {D.~Y.}\ \bibnamefont
  {Kononenko}}, \bibinfo {author} {\bibfnamefont {K.~V.}\ \bibnamefont
  {Yershov}}, \bibinfo {author} {\bibfnamefont {U.~K.}\ \bibnamefont
  {R{\"{o}}{\ss}ler}}, \bibinfo {author} {\bibfnamefont {A.~V.}\ \bibnamefont
  {Tomilo}}, \bibinfo {author} {\bibfnamefont {J.}~\bibnamefont {Fassbender}},
  \bibinfo {author} {\bibfnamefont {J.}~\bibnamefont {van~den Brink}}, \bibinfo
  {author} {\bibfnamefont {D.}~\bibnamefont {Makarov}},\ and\ \bibinfo {author}
  {\bibfnamefont {D.~D.}\ \bibnamefont {Sheka}},\ }\bibfield  {title} {\bibinfo
  {title} {Curvilinear one-dimensional antiferromagnets},\ }\href
  {https://doi.org/10.1021/acs.nanolett.0c03246} {\bibfield  {journal}
  {\bibinfo  {journal} {Nano Letters}\ }\textbf {\bibinfo {volume} {20}},\
  \bibinfo {pages} {8157} (\bibinfo {year} {2020})}\BibitemShut {NoStop}%
\bibitem [{\citenamefont {Borysenko}\ \emph {et~al.}(2022)\citenamefont
  {Borysenko}, \citenamefont {Sheka}, \citenamefont {Fassbender}, \citenamefont
  {van~den Brink}, \citenamefont {Makarov},\ and\ \citenamefont
  {Pylypovskyi}}]{Borysenko22}%
  \BibitemOpen
  \bibfield  {author} {\bibinfo {author} {\bibfnamefont {Y.~A.}\ \bibnamefont
  {Borysenko}}, \bibinfo {author} {\bibfnamefont {D.~D.}\ \bibnamefont
  {Sheka}}, \bibinfo {author} {\bibfnamefont {J.}~\bibnamefont {Fassbender}},
  \bibinfo {author} {\bibfnamefont {J.}~\bibnamefont {van~den Brink}}, \bibinfo
  {author} {\bibfnamefont {D.}~\bibnamefont {Makarov}},\ and\ \bibinfo {author}
  {\bibfnamefont {O.~V.}\ \bibnamefont {Pylypovskyi}},\ }\bibfield  {title}
  {\bibinfo {title} {Field-induced spin reorientation transitions in
  antiferromagnetic ring-shaped spin chains},\ }\href
  {https://doi.org/10.1103/physrevb.106.174426} {\bibfield  {journal} {\bibinfo
   {journal} {Physical Review B}\ }\textbf {\bibinfo {volume} {106}},\ \bibinfo
  {pages} {174426} (\bibinfo {year} {2022})}\BibitemShut {NoStop}%
\bibitem [{\citenamefont {Castillo-Sep{\'{u}}lveda}\ \emph
  {et~al.}(2017)\citenamefont {Castillo-Sep{\'{u}}lveda}, \citenamefont
  {Escobar}, \citenamefont {Altbir}, \citenamefont {Krizanac},\ and\
  \citenamefont {Vedmedenko}}]{Castillo-Sepulveda17}%
  \BibitemOpen
  \bibfield  {author} {\bibinfo {author} {\bibfnamefont {S.}~\bibnamefont
  {Castillo-Sep{\'{u}}lveda}}, \bibinfo {author} {\bibfnamefont {R.~A.}\
  \bibnamefont {Escobar}}, \bibinfo {author} {\bibfnamefont {D.}~\bibnamefont
  {Altbir}}, \bibinfo {author} {\bibfnamefont {M.}~\bibnamefont {Krizanac}},\
  and\ \bibinfo {author} {\bibfnamefont {E.~Y.}\ \bibnamefont {Vedmedenko}},\
  }\bibfield  {title} {\bibinfo {title} {Magnetic {M}{\"{o}}bius stripe without
  frustration: Noncollinear metastable states},\ }\href
  {https://doi.org/10.1103/physrevb.96.024426} {\bibfield  {journal} {\bibinfo
  {journal} {Physical Review B}\ }\textbf {\bibinfo {volume} {96}},\ \bibinfo
  {pages} {024426} (\bibinfo {year} {2017})}\BibitemShut {NoStop}%
\bibitem [{\citenamefont {Cador}\ \emph {et~al.}(2004)\citenamefont {Cador},
  \citenamefont {Gatteschi}, \citenamefont {Sessoli}, \citenamefont {Larsen},
  \citenamefont {Overgaard}, \citenamefont {Barra}, \citenamefont {Teat},
  \citenamefont {Timco},\ and\ \citenamefont {Winpenny}}]{Cador04}%
  \BibitemOpen
  \bibfield  {author} {\bibinfo {author} {\bibfnamefont {O.}~\bibnamefont
  {Cador}}, \bibinfo {author} {\bibfnamefont {D.}~\bibnamefont {Gatteschi}},
  \bibinfo {author} {\bibfnamefont {R.}~\bibnamefont {Sessoli}}, \bibinfo
  {author} {\bibfnamefont {F.~K.}\ \bibnamefont {Larsen}}, \bibinfo {author}
  {\bibfnamefont {J.}~\bibnamefont {Overgaard}}, \bibinfo {author}
  {\bibfnamefont {A.-L.}\ \bibnamefont {Barra}}, \bibinfo {author}
  {\bibfnamefont {S.~J.}\ \bibnamefont {Teat}}, \bibinfo {author}
  {\bibfnamefont {G.~A.}\ \bibnamefont {Timco}},\ and\ \bibinfo {author}
  {\bibfnamefont {R.~E.~P.}\ \bibnamefont {Winpenny}},\ }\bibfield  {title}
  {\bibinfo {title} {The magnetic {M}{\"{o}}bius strip: Synthesis, structure,
  and magnetic studies of odd-numbered antiferromagnetically coupled wheels},\
  }\href {https://doi.org/10.1002/anie.200460211} {\bibfield  {journal}
  {\bibinfo  {journal} {Angewandte Chemie International Edition}\ }\textbf
  {\bibinfo {volume} {43}},\ \bibinfo {pages} {5196} (\bibinfo {year}
  {2004})}\BibitemShut {NoStop}%
\bibitem [{\citenamefont {Baker}\ \emph {et~al.}(2012)\citenamefont {Baker},
  \citenamefont {Waldmann}, \citenamefont {Piligkos}, \citenamefont {Bircher},
  \citenamefont {Cador}, \citenamefont {Carretta}, \citenamefont {Collison},
  \citenamefont {Fernandez-Alonso}, \citenamefont {McInnes}, \citenamefont
  {Mutka}, \citenamefont {Podlesnyak}, \citenamefont {Tuna}, \citenamefont
  {Ochsenbein}, \citenamefont {Sessoli}, \citenamefont {Sieber}, \citenamefont
  {Timco}, \citenamefont {Weihe}, \citenamefont {G{\"{u}}del},\ and\
  \citenamefont {Winpenny}}]{Baker12}%
  \BibitemOpen
  \bibfield  {author} {\bibinfo {author} {\bibfnamefont {M.~L.}\ \bibnamefont
  {Baker}}, \bibinfo {author} {\bibfnamefont {O.}~\bibnamefont {Waldmann}},
  \bibinfo {author} {\bibfnamefont {S.}~\bibnamefont {Piligkos}}, \bibinfo
  {author} {\bibfnamefont {R.}~\bibnamefont {Bircher}}, \bibinfo {author}
  {\bibfnamefont {O.}~\bibnamefont {Cador}}, \bibinfo {author} {\bibfnamefont
  {S.}~\bibnamefont {Carretta}}, \bibinfo {author} {\bibfnamefont
  {D.}~\bibnamefont {Collison}}, \bibinfo {author} {\bibfnamefont
  {F.}~\bibnamefont {Fernandez-Alonso}}, \bibinfo {author} {\bibfnamefont
  {E.~J.~L.}\ \bibnamefont {McInnes}}, \bibinfo {author} {\bibfnamefont
  {H.}~\bibnamefont {Mutka}}, \bibinfo {author} {\bibfnamefont
  {A.}~\bibnamefont {Podlesnyak}}, \bibinfo {author} {\bibfnamefont
  {F.}~\bibnamefont {Tuna}}, \bibinfo {author} {\bibfnamefont {S.}~\bibnamefont
  {Ochsenbein}}, \bibinfo {author} {\bibfnamefont {R.}~\bibnamefont {Sessoli}},
  \bibinfo {author} {\bibfnamefont {A.}~\bibnamefont {Sieber}}, \bibinfo
  {author} {\bibfnamefont {G.~A.}\ \bibnamefont {Timco}}, \bibinfo {author}
  {\bibfnamefont {H.}~\bibnamefont {Weihe}}, \bibinfo {author} {\bibfnamefont
  {H.~U.}\ \bibnamefont {G{\"{u}}del}},\ and\ \bibinfo {author} {\bibfnamefont
  {R.~E.~P.}\ \bibnamefont {Winpenny}},\ }\bibfield  {title} {\bibinfo {title}
  {Inelastic neutron scattering studies on the odd-membered antiferromagnetic
  wheel {Cr}${}_{8}${Ni}},\ }\href {https://doi.org/10.1103/PhysRevB.86.064405}
  {\bibfield  {journal} {\bibinfo  {journal} {Physical Review B}\ }\textbf
  {\bibinfo {volume} {86}},\ \bibinfo {pages} {064405} (\bibinfo {year}
  {2012})}\BibitemShut {NoStop}%
\bibitem [{\citenamefont {O'Brien}\ \emph {et~al.}(2011)\citenamefont
  {O'Brien}, \citenamefont {Petit}, \citenamefont {Lewis}, \citenamefont
  {Cowburn}, \citenamefont {Read}, \citenamefont {Sampaio}, \citenamefont
  {Zeng},\ and\ \citenamefont {Jausovec}}]{OBrien11}%
  \BibitemOpen
  \bibfield  {author} {\bibinfo {author} {\bibfnamefont {L.}~\bibnamefont
  {O'Brien}}, \bibinfo {author} {\bibfnamefont {D.}~\bibnamefont {Petit}},
  \bibinfo {author} {\bibfnamefont {E.~R.}\ \bibnamefont {Lewis}}, \bibinfo
  {author} {\bibfnamefont {R.~P.}\ \bibnamefont {Cowburn}}, \bibinfo {author}
  {\bibfnamefont {D.~E.}\ \bibnamefont {Read}}, \bibinfo {author}
  {\bibfnamefont {J.}~\bibnamefont {Sampaio}}, \bibinfo {author} {\bibfnamefont
  {H.~T.}\ \bibnamefont {Zeng}},\ and\ \bibinfo {author} {\bibfnamefont
  {A.-V.}\ \bibnamefont {Jausovec}},\ }\bibfield  {title} {\bibinfo {title}
  {Tunable remote pinning of domain walls in magnetic nanowires},\ }\href
  {https://doi.org/10.1103/PhysRevLett.106.087204} {\bibfield  {journal}
  {\bibinfo  {journal} {Physical Review Letters}\ }\textbf {\bibinfo {volume}
  {106}},\ \bibinfo {pages} {087204} (\bibinfo {year} {2011})}\BibitemShut
  {NoStop}%
\bibitem [{\citenamefont {Donnelly}\ \emph {et~al.}(2022)\citenamefont
  {Donnelly}, \citenamefont {Hierro-Rodr\'{i}guez}, \citenamefont {Abert},
  \citenamefont {Witte}, \citenamefont {Skoric}, \citenamefont
  {Sanz-Hern\'{a}ndez}, \citenamefont {Finizio}, \citenamefont {Meng},
  \citenamefont {McVitie}, \citenamefont {Raabe}, \citenamefont {Suess},
  \citenamefont {Cowburn},\ and\ \citenamefont
  {Fern\'{a}ndez-Pacheco}}]{Donnelly22}%
  \BibitemOpen
  \bibfield  {author} {\bibinfo {author} {\bibfnamefont {C.}~\bibnamefont
  {Donnelly}}, \bibinfo {author} {\bibfnamefont {A.}~\bibnamefont
  {Hierro-Rodr\'{i}guez}}, \bibinfo {author} {\bibfnamefont {C.}~\bibnamefont
  {Abert}}, \bibinfo {author} {\bibfnamefont {K.}~\bibnamefont {Witte}},
  \bibinfo {author} {\bibfnamefont {L.}~\bibnamefont {Skoric}}, \bibinfo
  {author} {\bibfnamefont {D.}~\bibnamefont {Sanz-Hern\'{a}ndez}}, \bibinfo
  {author} {\bibfnamefont {S.}~\bibnamefont {Finizio}}, \bibinfo {author}
  {\bibfnamefont {F.}~\bibnamefont {Meng}}, \bibinfo {author} {\bibfnamefont
  {S.}~\bibnamefont {McVitie}}, \bibinfo {author} {\bibfnamefont
  {J.}~\bibnamefont {Raabe}}, \bibinfo {author} {\bibfnamefont
  {D.}~\bibnamefont {Suess}}, \bibinfo {author} {\bibfnamefont
  {R.}~\bibnamefont {Cowburn}},\ and\ \bibinfo {author} {\bibfnamefont
  {A.}~\bibnamefont {Fern\'{a}ndez-Pacheco}},\ }\bibfield  {title} {\bibinfo
  {title} {Complex free-space magnetic field textures induced by
  three-dimensional magnetic nanostructures},\ }\href
  {https://doi.org/10.1038/s41565-021-01027-7} {\bibfield  {journal} {\bibinfo
  {journal} {Nature Nanotechnology}\ }\textbf {\bibinfo {volume} {17}},\
  \bibinfo {pages} {136} (\bibinfo {year} {2022})}\BibitemShut {NoStop}%
\bibitem [{\citenamefont {Sheka}\ \emph
  {et~al.}(2015{\natexlab{b}})\citenamefont {Sheka}, \citenamefont {Kravchuk},\
  and\ \citenamefont {Gaididei}}]{Sheka15}%
  \BibitemOpen
  \bibfield  {author} {\bibinfo {author} {\bibfnamefont {D.~D.}\ \bibnamefont
  {Sheka}}, \bibinfo {author} {\bibfnamefont {V.~P.}\ \bibnamefont
  {Kravchuk}},\ and\ \bibinfo {author} {\bibfnamefont {Y.}~\bibnamefont
  {Gaididei}},\ }\bibfield  {title} {\bibinfo {title} {Curvature effects in
  statics and dynamics of low dimensional magnets},\ }\href
  {https://doi.org/10.1088/1751-8113/48/12/125202} {\bibfield  {journal}
  {\bibinfo  {journal} {Journal of Physics A: Mathematical and Theoretical}\
  }\textbf {\bibinfo {volume} {48}},\ \bibinfo {pages} {125202} (\bibinfo
  {year} {2015}{\natexlab{b}})}\BibitemShut {NoStop}%
\bibitem [{\citenamefont {Spaldin}\ \emph {et~al.}(2013)\citenamefont
  {Spaldin}, \citenamefont {Fechner}, \citenamefont {Bousquet}, \citenamefont
  {Balatsky},\ and\ \citenamefont {Nordstr\"{o}m}}]{Spaldin13}%
  \BibitemOpen
  \bibfield  {author} {\bibinfo {author} {\bibfnamefont {N.~A.}\ \bibnamefont
  {Spaldin}}, \bibinfo {author} {\bibfnamefont {M.}~\bibnamefont {Fechner}},
  \bibinfo {author} {\bibfnamefont {E.}~\bibnamefont {Bousquet}}, \bibinfo
  {author} {\bibfnamefont {A.}~\bibnamefont {Balatsky}},\ and\ \bibinfo
  {author} {\bibfnamefont {L.}~\bibnamefont {Nordstr\"{o}m}},\ }\bibfield
  {title} {\bibinfo {title} {Monopole-based formalism for the diagonal
  magnetoelectric response},\ }\href
  {https://doi.org/10.1103/physrevb.88.094429} {\bibfield  {journal} {\bibinfo
  {journal} {Physical Review B}\ }\textbf {\bibinfo {volume} {88}},\ \bibinfo
  {pages} {094429} (\bibinfo {year} {2013})}\BibitemShut {NoStop}%
\bibitem [{\citenamefont {Bhowal}\ and\ \citenamefont
  {Spaldin}(2022)}]{Bhowal22}%
  \BibitemOpen
  \bibfield  {author} {\bibinfo {author} {\bibfnamefont {S.}~\bibnamefont
  {Bhowal}}\ and\ \bibinfo {author} {\bibfnamefont {N.~A.}\ \bibnamefont
  {Spaldin}},\ }\bibfield  {title} {\bibinfo {title} {Magnetoelectric
  classification of skyrmions},\ }\href
  {https://doi.org/10.1103/physrevlett.128.227204} {\bibfield  {journal}
  {\bibinfo  {journal} {Physical Review Letters}\ }\textbf {\bibinfo {volume}
  {128}},\ \bibinfo {pages} {227204} (\bibinfo {year} {2022})}\BibitemShut
  {NoStop}%
\bibitem [{\citenamefont {Toulouse}(1977)}]{Toulouse77}%
  \BibitemOpen
  \bibfield  {author} {\bibinfo {author} {\bibfnamefont {G.}~\bibnamefont
  {Toulouse}},\ }\href@noop {} {\bibfield  {journal} {\bibinfo  {journal}
  {Commun. Phys.}\ }\textbf {\bibinfo {volume} {2}},\ \bibinfo {pages} {115}
  (\bibinfo {year} {1977})}\BibitemShut {NoStop}%
\bibitem [{\citenamefont {Toulouse}(1981)}]{Toulouse81}%
  \BibitemOpen
  \bibfield  {author} {\bibinfo {author} {\bibfnamefont {G.}~\bibnamefont
  {Toulouse}},\ }\bibinfo {title} {Spin glasses with special emphasis on
  frustration effects},\ in\ \href {https://doi.org/10.1007/bfb0012555} {\emph
  {\bibinfo {booktitle} {Lecture Notes in Physics}}}\ (\bibinfo  {publisher}
  {Springer Berlin Heidelberg},\ \bibinfo {address} {Berlin, Heidelberg},\
  \bibinfo {year} {1981})\ pp.\ \bibinfo {pages} {166--173}\BibitemShut
  {NoStop}%
\bibitem [{\citenamefont {Gibbs}\ \emph {et~al.}(2014)\citenamefont {Gibbs},
  \citenamefont {Mark}, \citenamefont {Lee}, \citenamefont {Eslami},
  \citenamefont {Schamel},\ and\ \citenamefont {Fischer}}]{Gibbs14}%
  \BibitemOpen
  \bibfield  {author} {\bibinfo {author} {\bibfnamefont {J.~G.}\ \bibnamefont
  {Gibbs}}, \bibinfo {author} {\bibfnamefont {A.~G.}\ \bibnamefont {Mark}},
  \bibinfo {author} {\bibfnamefont {T.-C.}\ \bibnamefont {Lee}}, \bibinfo
  {author} {\bibfnamefont {S.}~\bibnamefont {Eslami}}, \bibinfo {author}
  {\bibfnamefont {D.}~\bibnamefont {Schamel}},\ and\ \bibinfo {author}
  {\bibfnamefont {P.}~\bibnamefont {Fischer}},\ }\bibfield  {title} {\bibinfo
  {title} {Nanohelices by shadow growth},\ }\href
  {https://doi.org/10.1039/c4nr00403e} {\bibfield  {journal} {\bibinfo
  {journal} {Nanoscale}\ }\textbf {\bibinfo {volume} {6}},\ \bibinfo {pages}
  {9457} (\bibinfo {year} {2014})}\BibitemShut {NoStop}%
\bibitem [{\citenamefont {Monterde}(2024)}]{Monterde24}%
  \BibitemOpen
  \bibfield  {author} {\bibinfo {author} {\bibfnamefont {J.}~\bibnamefont
  {Monterde}},\ }\bibfield  {title} {\bibinfo {title} {Salkowski curves and
  spherical epicycloids},\ }\href {https://doi.org/10.1007/s00022-024-00730-9}
  {\bibfield  {journal} {\bibinfo  {journal} {Journal of Geometry}\ }\textbf
  {\bibinfo {volume} {115}},\ \bibinfo {pages} {31} (\bibinfo {year}
  {2024})}\BibitemShut {NoStop}%
\bibitem [{hzd()}]{hzdrcluster}%
  \BibitemOpen
  \href {http://www.hzdr.de} {\bibinfo {title} {{High Performance Computing at
  Helmholtz--Zentrum Dresden--Rossendorf}}},\ \bibinfo {howpublished}
  {\url{http://www.hzdr.de}}\BibitemShut {NoStop}%
\bibitem [{sla()}]{slasi}%
  \BibitemOpen
  \href {http://slasi.knu.ua} {\bibinfo {title} {\textsf{SLaSi} spin--lattice
  simulations package}}\BibitemShut {NoStop}%
\bibitem [{\citenamefont {Hu}\ \emph {et~al.}(2011)\citenamefont {Hu},
  \citenamefont {Lundgren},\ and\ \citenamefont {Niemi}}]{Hu11}%
  \BibitemOpen
  \bibfield  {author} {\bibinfo {author} {\bibfnamefont {S.}~\bibnamefont
  {Hu}}, \bibinfo {author} {\bibfnamefont {M.}~\bibnamefont {Lundgren}},\ and\
  \bibinfo {author} {\bibfnamefont {A.~J.}\ \bibnamefont {Niemi}},\ }\bibfield
  {title} {\bibinfo {title} {Discrete {F}renet frame, inflection point
  solitons, and curve visualization with applications to folded proteins},\
  }\href {https://doi.org/10.1103/physreve.83.061908} {\bibfield  {journal}
  {\bibinfo  {journal} {Physical Review E}\ }\textbf {\bibinfo {volume} {83}},\
  \bibinfo {pages} {061908} (\bibinfo {year} {2011})}\BibitemShut {NoStop}%
\end{thebibliography}
%

\end{document}